\documentclass[Afour,sageh,times]{sagej}

\usepackage[utf8]{inputenc}
\DeclareUnicodeCharacter{2003}{"}

\usepackage{amsmath}
\usepackage{amssymb}

\usepackage{algorithm}
\usepackage{algpseudocode}
\usepackage[dvipsnames]{xcolor}
\usepackage{makecell}
\usepackage{physics}
\usepackage{subcaption}
\usepackage{url}

\allowdisplaybreaks % Page break of equations

\newcommand{\vc}[1]{\mbox{\boldmath$#1$\unboldmath}}
\newcommand{\dt}{\Delta t}

\newcommand{\dx}{\Delta x}

\newcommand{\dz}{\Delta z}
\newcommand{\eb}{\vc{e}}

\newcommand{\Ib}{\vc{I}}
\newcommand{\kb}{\vc{k}}

\newcommand{\qb}{\vc{q}}

\newcommand{\rb}{\vc{r}}

\newcommand{\ub}{\vc{u}}

\newcommand{\xb}{\vc{x}}
\newcommand{\xib}{\vc{\xi}}

\newcommand{\intO}[1]{\int_{\Omega}#1d\Omega}
\newcommand{\intOe}[1]{\int_{\Omega_e}#1d\Omega_e}

\usepackage{moreverb,url}

\usepackage[colorlinks,bookmarksopen,bookmarksnumbered,citecolor=red,urlcolor=red]{hyperref}

\newcommand\BibTeX{{\rmfamily B\kern-.05em \textsc{i\kern-.025em b}\kern-.08em
T\kern-.1667em\lower.7ex\hbox{E}\kern-.125emX}}

\def\volumeyear{2025}
\begin{document}

\runninghead{Kang, Giraldo, and Camp }

\title{A GPU-accelerated simulation of rapid intensification of a tropical cyclone with observed heating}

\author{Soonpil Kang\affilnum{1,2}, Francis X. Giraldo\affilnum{1}, and Seth Camp\affilnum{3}}

\affiliation{\affilnum{1}Department of Applied Mathematics, Naval Postgraduate School, Monterey, CA, USA \\
\affilnum{2} Atmospheric, Earth, and Energy Division, Lawrence Livermore National Laboratory, Livermore, CA, USA \\
\affilnum{3} NVIDIA}

% \affiliation{\affilnum{2}NVIDIA}

\corrauth{Soonpil Kang}

\email{kang18@llnl.gov}

\begin{abstract}
  This paper presents a limited-area atmospheric simulation of a tropical cyclone accelerated using GPUs. The OpenACC directive-based programming model is used to port the atmospheric model to the GPU. The GPU implementation of the main functions and kernels is discussed. The GPU-accelerated code produces high-fidelity simulations of a realistic tropical cyclone forced by observational latent heating. Performance tests show that the GPU-accelerated code yields energy-efficient simulations and scales well in both the strong and weak limit.
\end{abstract}


\maketitle

% !TEX root = ../main.tex
\section{Introduction}
\label{sec:introduction}

Tropical cyclones (TCs) are a natural hazard that significantly impact human life and economies. Despite their huge impact, TC evolution is a physical process that is poorly understood currently. In particular, the prediction of TC intensity has been a challenging problem because of complexities in the multiscale interactions of convective-scale processes and synoptic-scale environment \citep{fischer2019climatological}. This is true especially when TCs undergo rapid intensification (RI), where TCs drastically intensify in a short period of time -- typically change in wind speed of around 30 knots within 24 hours. To explore and study the RI mechanisms, we perform high-fidelity TC simulations. Modeling and numerical simulations of TCs can enhance our understanding of the underlying geophysical processes and may improve the accuracy of forecasts and risk assessment. However, one of the main challenges in the modeling of TC evolution is the high computational cost to resolve the fine-scale features needed for accurate prediction of TC intensity. To achieve computationally efficient simulations, we port the limited-area atmospheric model to the graphics processing unit (GPU) using a multi-CPU multi-GPU approach.

A nonhydrostatic atmospheric model is used to simulate TC evolutions. The code we use to perform these simulations is called xNUMA, a new version of the NUMA (Nonhydrostatic Unified Model of the Atmosphere) code \citep{kelly2012continuous,giraldo2013implicit}. The xNUMA code is designed specifically for a multiscale modeling framework (MMF) for atmospheric flows, and is written in modern Fortran using an object-oriented approach. Because all the spatial discretization and time-integration methods are encapsulated in the form of objects, it is capable of instantiating multiple simulations simultaneously during execution.

Previously, the NUMA nonhydrostatic atmospheic model, which employs element-based Galerkin methods, was ported to GPUs \citep{abdi2019acceleration, abdi2019gpu} using OCCA \citep{medina2014occa}. Various time integration methods, including explicit and semi-implicit schemes, were ported and demonstrated good scalability for numerical weather problems. However, using a hardware-agnostic application programming interface (API) like OCCA increases development complexity. It requires writing an entirely separate parallel branch of code in different languages, as well as modifying data structures and kernel execution patterns. In this work, we explore a hardware-specific directive-based approach to simplify offloading while maintaining performance.

In porting xNUMA to GPUs, we chose to approach the problem using a directive based implementation. The main advantage of directive-based approaches is the ability to easily make large GPU coding contributions to the code base while still maintaining a correct and performant CPU code base. In this work, we ported xNUMA to GPUs using the OpenACC programming standard \citep{openaccspec} as implemented in the NVIDIA compiler \citep{NVIDIAHPCCompilersRefGuide}.

High-order Galerkin methods are attractive for hurricane research due to their low numerical dissipation. An earlier study \citep{guimond2016impacts} reveals that high-order methods better capture the vortex response to asymmetric thermal perturbations in hurricane RI due to reduced damping of heat and kinetic energies. Recently, an adaptive mesh refinement technique has been incorporated into the high-order Galerkin method to efficiently capture the fine features of hurricanes \citep{tissaoui2024accelerating}.

High-order Galerkin methods have been shown to be performant on GPUs and well-suited for modern high-performance computing architectures. The Center for Efficient Exascale Discretizations (CEED) project documented applications of these methods in \citep{kolev2021efficient}. The finite element solver MFEM has been ported to GPUs \citep{vargas2022matrix,andrej2024high} by integrating the RAJA abstraction layer, which facilitates access to various backend programming models for parallelization of loop kernels. The spectral element Navier-Stokes solver Nek5000 has been ported to GPUs using OpenACC \citep{otero2019openacc} and OCCA \citep{fischer2022nekrs}. These works commonly adopted a matrix-free approach for GPU acceleration and reported performance improvements as the polynomial order of the basis functions increased.

The remainder of the paper is organized as follows. Section \ref{sec:atmosphere} describes the nonhydrostatic atmospheric model. In this section, the governing equations, spatial, and temporal discretizations are summarized. Section \ref{sec:implementation} presents the porting and optimization of the code for single and multiple GPU systems. Section \ref{sec:setup} presents the numerical setup for tropical cyclone simulations. Section \ref{sec:results} shows the performance test results. The conclusions are drawn in Section \ref{sec:conclusion}.

\section{Nonhydrostatic atmospheric model}\label{sec:atmosphere}
% !TEX root = ../main.tex

\subsection{Governing equations}
\label{sec:governing_equations}

Let us consider a fixed spatial domain $\Omega$ and a time interval $(0,t_f]$. The governing equations for the nonhydrostatic model of the atmosphere can be written as follows:
%\begin{subequations}\label{eq:governing}
%	\begin{align}
%		&\pdv{\rho'}{t} + \div\qty\big((\rho_0+\rho')\ub) = 0, \label{eq:governing:rho}
%  \\
%	&\pdv{\ub}{t} + \ub\cdot\grad\ub + \dfrac{1}{\rho_0+\rho'} \grad p' + \dfrac{\rho'}{\rho_0+\rho'} g\kb  = \nu\grad^2\ub, \label{eq:governing:u}
%  \\
%	&\pdv{\theta'}{t} + \ub\cdot\grad(\theta_0+\theta') = S_{\theta} + \nu\grad^2\theta', \label{eq:governing:theta}
%	\end{align}
%\end{subequations}
\begin{subequations}\label{eq:governing}
	\begin{align}
		&\pdv{\rho'}{t} + \div\qty\big((\rho_0+\rho')\ub) = 0, \label{eq:governing:rho}
  \\
	&\pdv{\ub}{t} + \ub\cdot\grad\ub + \dfrac{1}{\rho_0+\rho'} \grad p' + \dfrac{\rho'}{\rho_0+\rho'} g\kb  = \mathbf{H}_{\ub}, \label{eq:governing:u}
  \\
	&\pdv{\theta'}{t} + \ub\cdot\grad(\theta_0+\theta') = S_{\theta} + {H}_{\theta}, \label{eq:governing:theta}
	\end{align}
\end{subequations}
where $\rho_0$ is the reference density, $\rho'$ is the density perturbation, $\ub$ is the velocity vector, $\theta_0$ is the reference potential temperature, $\theta'$ represents the potential temperature perturbation, $p$ is the pressure, $g$ is the gravitational acceleration, $\kb$ is the upward-pointing unit vector, $S_{\theta}$ is the heat source, and $\mathbf{H}_{\mathbf{q}}$ is the hyper-diffusion operator for the variable $\mathbf{q}$, which we define in Section \ref{sec:hyperdiffusion}.
%and $\nu$ is the kinematic viscosity.
Equations \eqref{eq:governing:rho}--\eqref{eq:governing:theta} represent local mass conservation, momentum balance, and thermodynamics, respectively. We assume that the reference fields are in hydrostatic balance and dependent only on the vertical coordinate $z$, i.e., $\dv{p_0}{z}=-\rho_0 g$. 

The prognostic variables for the system \eqref{eq:governing} are $\qb=(\rho', \ub, \theta')^\mathcal{T}$, where the superscript $\mathcal{T}$ denotes the transpose operator. Eq.~\eqref{eq:governing} can be rewritten more concisely in the compact form as
\begin{equation}\label{eq:governing:compact}
	\pdv{\qb}{t} = \mathcal{S}(\qb),
\end{equation}
where $\mathcal{S}(\qb)$ contains all terms in the equations apart from the time derivatives. 

The pressure of dry air is calculated using the thermodynamic equation of state,
\begin{equation}
  p = p_a \qty(\frac{\rho R \theta}{p_a})^{\gamma}
\end{equation}
where $p_a$ is the atmospheric pressure at the ground, $R$ is the specific gas constant of dry air, and $\gamma=c_p/c_v$ with the specific heat at constant pressure $c_p$ and at constant volume $c_v$.

\subsection{Element-based Galerkin method}
\label{sec:ebg}

The element-based Galerkin method decomposes the spatial domain $\Omega\subset\mathbb{R}^d$ ($d=$2 or 3) into $N_e$ disjoint elements $\Omega_e$ such as $\Omega = \bigcup_{e=1}^{N_e}\Omega_e$. In our model, the $\Omega_e$ are either quadrilaterals (in 2D) or hexahedra (in 3D). Within each element $\Omega_e$, the prognostic vector $\qb$ is approximated as a finite-dimensional projection $\qb_N$ using basis functions $\psi_j(\xb)$ and element-wise nodal coefficients $\qb_j^{(e)}(t)$ such that
\begin{equation}\label{eq:approximation_q}
  \qb_N(\xb,t) = \sum_{j=1}^{M_N} \psi_j(\xb) \qb_j^{(e)}(t),
\end{equation}
where $N$ denotes the order of the basis functions, $M_N=(N+1)^d$ is the number of nodes per element, and the superscript $(e)$ denotes element-wise quantities. The basis functions are constructed as tensor products of $N$-th order Lagrange polynomials $h_\alpha$ associated with the Legendre-Gauss-Lobatto (LGL) points, given by
\begin{equation}
  \psi_i(\xi,\eta,\zeta) = h_\alpha(\xi)\otimes h_\beta(\eta) \otimes h_\gamma(\zeta),
\end{equation}
where $\alpha,\beta,\gamma\in \qty{1,\cdots,N+1}$, and $(\xi,\eta,\zeta)$ are the element coordinates in the reference domain $E=[-1,1]^d$ that are mapped from the physical coordinate $\xb$ via metric terms. 

For a continuous Galerkin (CG) method, let us consider a finite-dimensional Sobolev space $\mathcal{V_N}$ defined as
\begin{equation}
  \mathcal{V_N} = \qty\big{ \psi \; | \; \psi\in H^1(\Omega) \text{ and } \psi\in\mathcal{P}_N \left( \Omega_e \right)},
\end{equation}
where $\mathcal{P}_N \left( \Omega_e \right)$ is the set of all polynomials of degree less than or equal to $N$ on the element $\Omega_e$. The discrete weak form of the governing equation in \eqref{eq:governing:compact} is obtained by multiplying it by a test function and integrating over the domain, which is stated as follows: find $\qb_N\in\mathcal{V_N}$ such that for all $\psi\in\mathcal{V_N}$,
\begin{equation}\label{eq:governing:weak}
  \sum_{e=1}^{N_e}\intOe{\psi_i\pdv{\qb_N}{t}} = \sum_{e=1}^{N_e}\intOe{\psi_i \mathcal{S}(\qb_N)}.
\end{equation}
In 3D, the integral over the domain $\Omega$ is computed using a quadrature rule based on the LGL points as follows:
\begin{align}\label{eq:integral}
  \intO{(\,\cdot\,)} &= \sum_{e=1}^{N_e}\intOe{(\,\cdot\,)} \notag\\
  &= \sum_{e=1}^{N_e}\sum_{i=1}^{N_{\xi}+1}\sum_{j=1}^{N_{\eta}+1}\sum_{k=1}^{N_{\zeta}+1}w^{(e)}_{ijk} J^{(e)}_{ijk}(\,\cdot\,),
\end{align}
where $N_{\xi}$, $N_{\eta}$, and $N_{\zeta}$ are the orders of the polynomial basis functions in each direction, and $w^{(e)}_{ijk}$ and $J^{(e)}_{ijk}$ denote the weight and Jacobian determinant associated with the LGL points, respectively.

The discrete weak form in \eqref{eq:governing:weak} yields a matrix-vector form written as
\begin{equation}
  M_{IJ}\dv{\qb_J}{t} = R_I(\qb_N),
\end{equation}
where $I,J\in\qty{1,\cdots,N_p}$ and $N_p$ is the total number of global grid points, and the standard summation convention is assumed unless otherwise specified. The global mass matrix $M_{IJ}$ and the right-hand side (RHS) vector $R_I$ are constructed by applying the global assembly or direct stiffness summation (DSS) operator $\bigwedge_{e=1}^{N_e}$ to the element-wise mass matrices and RHS vectors as follows:
\begin{align}
  M_{IJ} &= \bigwedge_{e=1}^{N_e} \intOe{\psi_i (\xb) \psi_j (\xb)}, \\
  R_{I}(\qb_N) &= \bigwedge_{e=1}^{N_e} \intOe{\psi_i (\xb) \mathcal{S}(\qb_N)}, \label{eq:RHSvec}
\end{align}
where the DSS operator performs summation through mapping from local indices to global indices as $(i,e)\rightarrow I$ and $(j,e)\rightarrow J$. We choose the integration points to be co-located with the interpolation points (inexact integration), which yields accurate numerical integration for $N\ge4$ \citep{giraldo1998lagrange}. This choice makes the mass matrix diagonal, i.e., $M_{I}$, due to the cardinal property of the basis functions, simplifying inversion. Consequently, we obtain
\begin{equation}\label{eq:governing:matrix}
  \dv{\qb_I}{t} = M^{-1}_{I}R_I(\qb_N).
\end{equation}
Eq.~\eqref{eq:governing:matrix} is used to calculate the numerical solutions in the simulations.

% !TEX root = ../main.tex

\subsection{Implicit-explicit (IMEX) time integration}
\label{sec:time_integration}

We integrate the discretized equation \eqref{eq:governing:matrix} in time using an implicit-explicit (IMEX) time integrator \citep{giraldo2013implicit}, which splits the operator $\mathcal{S}(\qb)$ in the governing equations as follows:
\begin{equation}
    \pdv{\qb}{t} = \qty\big[\mathcal{S}(\qb) - \mathcal{L}(\qb)] + \mathcal{L}(\qb).
\end{equation}
The terms in the square bracket account for slower advective waves and are discretized explicitly. The remaining terms in the linear operator $\mathcal{L}(\qb)$ represent fast waves, such as acoustic and gravity waves, and are discretized implicitly. This IMEX strategy enables the use of larger time steps by eliminating the tight CFL restriction due to the fast waves. In this work, we employ the second-order additive Runge-Kutta (ARK2) scheme proposed in \citep{giraldo2013implicit}, which belongs to a family of linear multistage schemes. This time integrator is also used in \citep{giraldo2024performance} referred to as ARK(2,3,2)b. 

We solve the linear system that arises from the implicit part of the IMEX formulation using the iterative Krylov subspace GMRES \citep{saad1986gmres} in a fully matrix-free fashion. In this approach, the linearized operator is computed on the fly with no information precomputed or stored. The only memory required is to store the Krylov vectors, which is typically fewer than 20 in the simulations presented.

The complexity of the algorithm is analyzed in our earlier work. Detailed information can be found in
our previous paper \citep{kang2024multiscale}.

% !TEX root = ../main.tex
\subsection{Tensor-based Hyper-Diffusion}\label{sec:hyperdiffusion}

We introduce numerical hyper-diffusion by adding the following term to the right-hand side of the continuous governing equations
\begin{equation}\label{eq:hyperdiffusion}
%\dv{\qb}{t} = \mathcal{S}(\qb) +  (-1)^{\alpha+1}(\nabla\cdot\vc{\tau}\nabla)^\alpha \qb,
\mathbf{H}_{\qb} = (-1)^{\alpha+1}(\nabla\cdot\vc{\tau}\nabla)^\alpha \qb,
\end{equation}
%where $\mathcal{S}(\qb)$ represents the spatial operators of the compressible Euler equations, 
where $\qb$ is the state variable, $\vc{\tau}$  the viscosity tensor, and $\alpha$ is the order. In this work, we employed $\alpha=2$, which yields fourth order hyper-diffusion. In this formulation, we adopt the tensor form of viscosity proposed in \citep{guba2014spectral}, but modify the original formulation by multiplying the viscosity tensor $\vc{\tau}$ by each application of the Laplacian operator. This modification benefits the memory storage of metric terms. The viscosity tensor $\vc{\tau}$ is defined through the eigendecomposition as follows:
\begin{equation}\label{eq:tau_tensor}
\vc{\tau}=\vc{JE} \begin{pmatrix}
\nu_1\lambda_1^{-1} & 0 & 0 \\ 0 & \nu_2\lambda_2^{-1} & 0 \\ 0 & 0 & \nu_3\lambda_3^{-1}
\end{pmatrix}(\vc{JE})^{\mathcal{T}},
\end{equation}
where $\vc{J}=\pdv{\xb}{\xib}$ is the Jacobian matrix of the isoparametric mapping, $\left( \nu_1,\nu_2,\nu_3 \right)$ are the viscosity parameters that act in the principal axes of the metric tensor $\vc{G}=\vc{J}^{-1}\vc{J}^{-\mathcal{T}}$, and $\left( \lambda_1^{-1},\lambda_2^{-1},\lambda_3^{-1} \right)$ are the eigenvalues of $\vc{G}$ in ascending order ($\lambda_1>\lambda_2>\lambda_3$). The matrix $\vc{E}$ contains the normalized eigenvectors $\left( \vc{e}_1, \vc{e}_2, \vc{e}_3 \right)$ of $\vc{G}$ in its columns, i.e., $\vc{E}=\left[\vc{e}_1 | \vc{e}_2 | \vc{e}_3\right]$. $\vc{E}$ is orthonormal because $\vc{G}$ is symmetric. The viscosity parameters $\nu_i$ have physical dimensions $L^2/T^{1/\alpha}$, while kinematic (physical) hyperviscosity $\nu_p$  has dimensions $L^{2\alpha}/T$.
The viscosity parameters can be scaled by the element size $\Delta x$ and time-step $\Delta t$ as follows
\begin{equation}
\label{eq:viscosity_scaled}
\nu_i \equiv (c_i)^{1/\alpha}\frac{(\Delta x_i)^2}{(\Delta t)^{1/\alpha}}  \qquad (i=1,\ldots,3)
\end{equation}
where $c_i=\left( c_1,c_2,c_3 \right)$ are the dimensionless coefficients that independently control the amount of numerical hyper-diffusion in the principal directions. Further details on our hyper-diffusion model are provided in our paper \citep{giraldo2024performance}.

%------------------------------------------------------%
%   								Complexity
%------------------------------------------------------%

% \input{sections/Complexity.tex}

%------------------------------------------------------%
%   								Implementation
%------------------------------------------------------%

\section{GPU implementation}\label{sec:implementation}
\subsection{Base code}\label{sec:basecode}

The xNUMA code is written in modern Fortran and the base CPU code is parallelized using a Message Passing Interface (MPI) implementation. The simulator utilizes the dynamical core of the nonhydrostatic unified model of the atmosphere, referred to as NUMA \citep{kelly2012continuous,giraldo2013implicit}. Each simulator object contains objects for input, spatial discretization, time integration, and solvers. Consequently, all the data and functions that constitute a simulation are members of the simulator objects.

\begin{figure}[h!]
  \setlength{\fboxsep}{0pt}%
  \setlength{\fboxrule}{0pt}%  
  \begin{center}
  \includegraphics[width=0.45\textwidth]{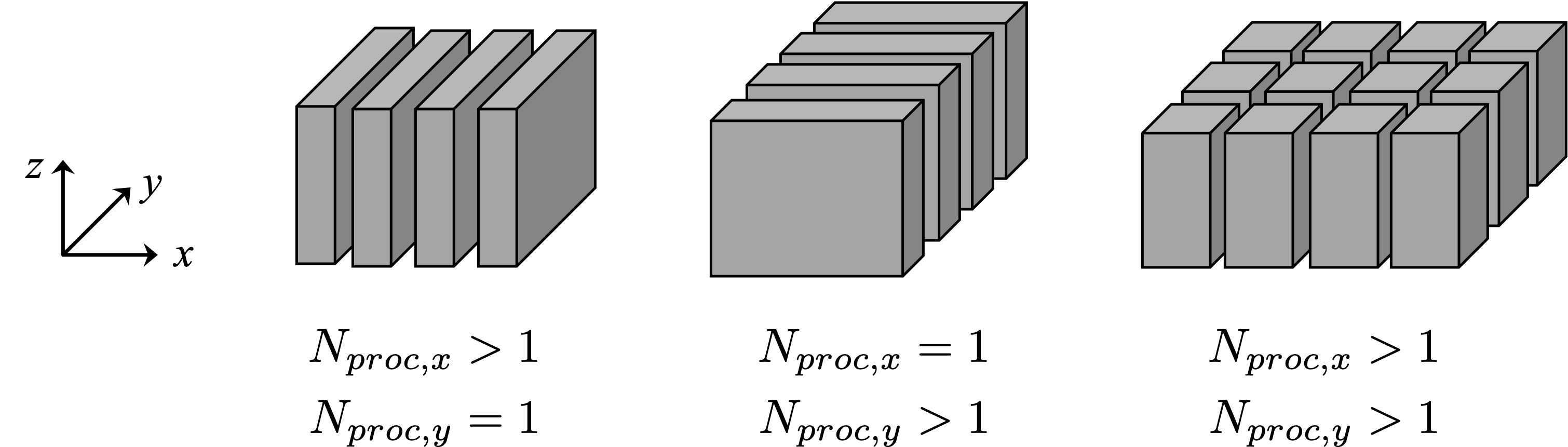}
  \end{center}
  \caption{Layouts of domain decomposition for limited-area models. ($N_{proc,x}$ and $N_{proc,y}$ are the number of processors in the $x$ and $y$ directions.)}\label{fig:parallel:decomposition}
\end{figure}

The simulator is implemented via distributed memory parallelism using MPI. We subdivide the computational domain for limited-area simulations into multiple subdomains of equal size using non-overlapping element-based partitioning. In this configuration, each element is exclusively assigned to a single partition, and each partition is assigned to an MPI process (or core). As a result, workload is well balanced among MPI processes. Figure \ref{fig:parallel:decomposition} illustrates three types of layouts for domain decomposition. We chose to keep the grid columns within a partition for a potential use of column-based microphysics model. The inter-process communications are required for the global DSS operation over the partitions. The partitioning of a grid is similar to what we implemented for the standard model. Details about the computational stencil for the element-based Galerkin method and the global DSS process can be found in \citep{kelly2012continuous}.

% !TEX root = ../main.tex
\subsection{OpenACC implementation}

\begin{figure}[h!]
  \setlength{\fboxsep}{0pt}%
  \setlength{\fboxrule}{0pt}%  
  \begin{center}
  \includegraphics[width=0.45\textwidth]{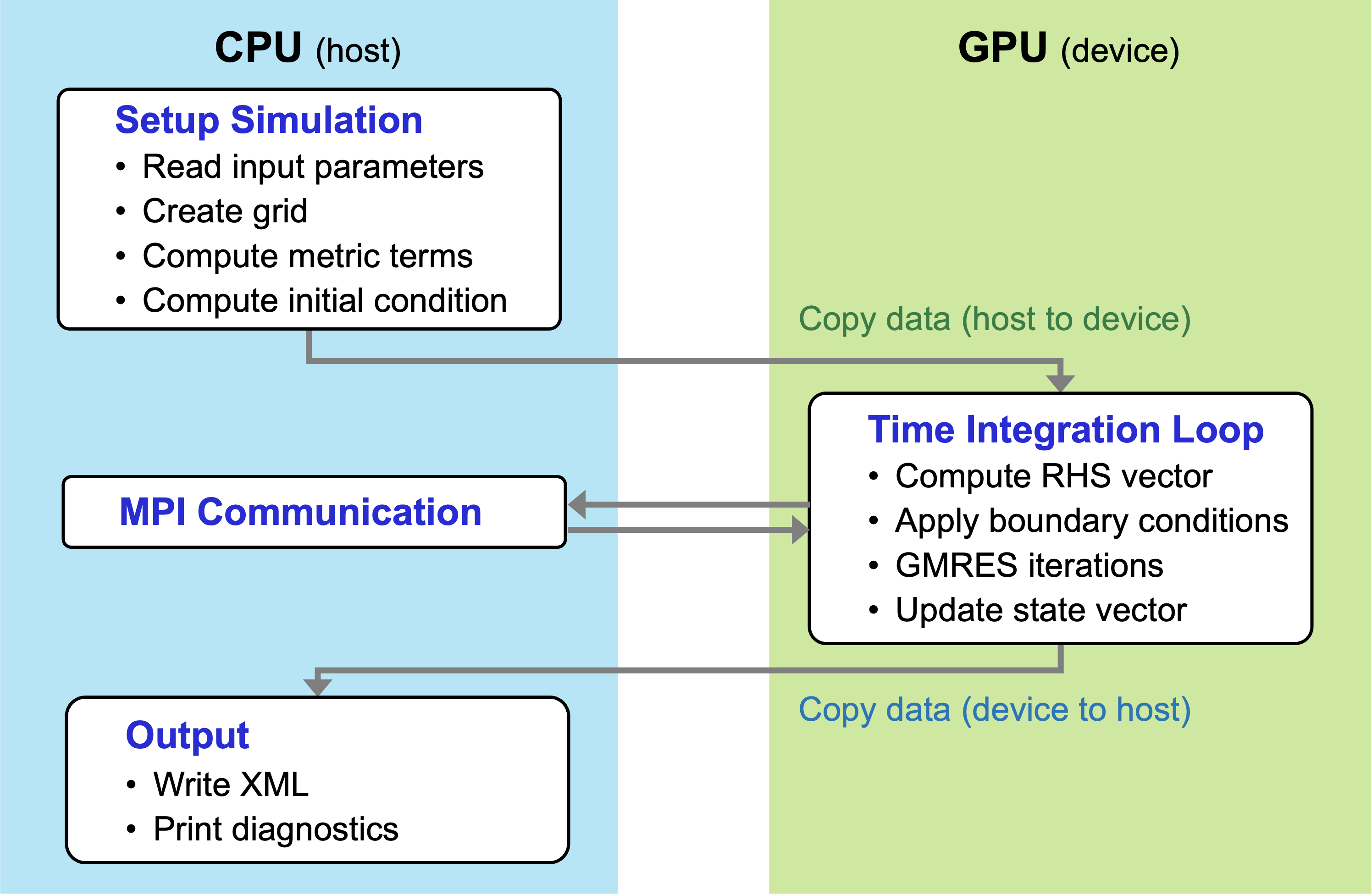}
  \end{center}
  \caption{Simulation workflow on the CPU and GPU.}\label{fig:porting}
\end{figure}

In this work, we ported the time integration task, a main component of the simulation, to GPUs using OpenACC directives. Figure \ref{fig:porting} illustrates the simulation workflow on the CPU and GPU. Initially, the simulation is set up on the CPUs by reading input parameters, creating the grid, and computing the intial conditions. Subsequently, all data is transferred to GPUs using OpenACC directives to copy data from the CPU to the GPU. On the GPUs, the simulation advances in time by calculating the RHS vector, performing GMRES iterations, and updating the state vector. To output XML outputs and diagnostics, we transfer the GPU value for the state vector to the CPU, again using OpenACC directives. For this work, we have included all data management directives explicitly rather than using NVIDIA compiler implemented unified or managed memory approaches. These latter memory approaches can simplify porting, but can require either code refactoring (managed memory) or bleeding edge software/hardware (unified memory) to function properly. Explicit data management gives us minimal code change requirements while providing the ability to maximize performance.

For communication across GPUs, we utilize OpenACC interoperability features that enable MPI communication directly between GPUs. The \texttt{host\_data} construct makes GPU data addresses accessible on the CPU, and arrays listed in the \texttt{use\_device} clause can be passed to CPU functions. If the hardware does not support CUDA-aware MPI or GPUDirect, standard network communication methods, such as Transmission Control Protocol (TCP), will be used.

%------------------------------------------------------%
% RHS kernel
%------------------------------------------------------%

\subsection{RHS function}

Algorithm \ref{alg:rhs} presents pseudocode for the function that computes the RHS vector, ported to the GPU using OpenACC directives. In the code, $N_e$ is the number of elements, while $N_{\xi}$, $N_{\eta}$, and $N_{\zeta}$ are the number of points within an element in each respective direction. $N_{\text{var}}$ denotes the number of variables at each node. The function $intma$ maps local nodal indices to global nodal indices. The RHS function consists of four kernels: global-to-local, gradient, divergence, and local-to-global kernels. Each kernel is parallelized using the \texttt{parallel loop} directive and optimized with gang and vector elements, which represent coarse-grain and fine-grain levels of parallelism, respectively. Vector parallelism operates within each gang in single instruction multiple threads (SIMT) lanes. The quadruple-nested loops over spectral elements and their nodal points are collapsed into a single loop using the \texttt{collapse} clause, which allows the compiler to maximize parallelism by mapping the nested loop to an unnested parallelizable loop with $N_e\times N_{\zeta}\times N_{\eta}\times N_{\xi}$ independent iterations. Despite also being a parallelizable loop, we found that allowing the innermost loop to be run sequentially maximized performance -- possibly due to memory or register access constraints.

\begin{algorithm}[htb]
  \scriptsize
  \caption{RHS function.}
  \label{alg:rhs}
  \begin{algorithmic}

  \Function{Compute\underline{\hspace{0.15cm}}RHS}{}

  \State $\color{magenta}!\$acc \; data \; present(q,q_e,f,rhs)$
  \State

  \State \textcolor{ForestGreen}{!Kernel: Global to Local}
  
  \State \textcolor{magenta}{$!\$acc \; parallel \; loop \; gang \; vector \; collapse(4)$}
  \For{$e=1:N_e, k=1:N_{\zeta}, j=1:N_{\eta}, i=1:N_{\xi}$}
  \State $I=intma(i,j,k,e)$
  \State \textcolor{magenta}{$!\$acc \; loop \; seq$}
  \For{$m=1:N_{\text{var}}$}
    \State $q_e(i,j,k,e,m)=q(m,I)$ 
  \EndFor
  \EndFor

  \State

  \State \textcolor{ForestGreen}{!Kernel: Gradient}
  \State Compute local gradient operator 

  \State

  \State \textcolor{ForestGreen}{!Kernel: Divergence}
  \State Compute local divergence operator  

  \State

  \State \textcolor{ForestGreen}{!Kernel: Local to Global}
  \State \textcolor{magenta}{$!\$acc \; parallel \; loop \; gang \; vector \; collapse(4)$}
  \For{$e=1:N_e, k=1:N_{\zeta}, j=1:N_{\eta}, i=1:N_{\xi}$}
  \State \textcolor{magenta}{$!\$acc \; loop \; seq$}
  \For{$m=1:N_{\text{var}}$}    
    \State Evaluate $f\qty(\nabla q(i,j,k,e,m))$
    \EndFor
    \State $I=intma(i,j,k,e)$
    \State \textcolor{magenta}{$!\$acc \; loop \; seq$}
    \For{$m=1:N_{\text{var}}$}    
      \State $rhs(m,I)=f\qty(\nabla q(i,j,k,e,m))$ 
    \EndFor
  \EndFor %m,i,j,k,e

  \State
  \State $\color{magenta}!\$acc \; end \; data$

  \EndFunction

  \end{algorithmic}
\end{algorithm}

%------------------------------------------------------%
% Gradient kernel
%------------------------------------------------------%

\subsection{Gradient kernel}

Algorithm \ref{alg:gradient} presents pseudocode for the function that computes the gradient fields of the prognostic variables. This function is the most computationally intensive part of the code and is executed when evaluating the right-hand side vector or the linearized operator. In the code, $q_{\xi}$, $q_{\eta}$, and $q_{\zeta}$ represent the differentiation of the state vector with respect to the reference coordinates, $\partial q/\partial\xi$, $\partial q/ \partial\eta$, and $\partial q/ \partial\zeta$, respectively. 

\begin{algorithm}[htb]
  \scriptsize
  \caption{Gradient kernel.}
  \label{alg:gradient}
  \begin{algorithmic}
  \Function{Compute\underline{\hspace{0.15cm}}Gradient}{}

  \State $\color{magenta}!\$acc \; data \; present(q_{\xi},q_{\eta},q_{\zeta})$
  \State
  
  \State \textcolor{magenta}{$!\$acc \; parallel \; loop \; gang \; vector \; collapse(4)$}
  \For{$e=1:N_e, k=1:N_{\zeta}, j=1:N_{\eta}, i=1:N_{\xi}$}
    \For{$m=1:N_{\text{var}}$}
      \State $hold1=0.0$
      \State $hold2=0.0$
      \State $hold3=0.0$
      \State \textcolor{magenta}{$!\$acc \; loop \; reduction(+:hold1,hold2,hold3)$}
      \For{$l=1:N_{\xi}$}
        \State $hold1 = hold1 + \pdv{\phi}{\xi} \qty(l,i) \times q(l,j,k,e,m)$
        \State $hold2 = hold2 + \pdv{\phi}{\eta} \qty(l,j) \times q(i,l,k,e,m)$
        \State $hold3 = hold3 + \pdv{\phi}{\zeta} \qty(l,k) \times q(i,j,l,e,m)$
      \EndFor
      \State $q_{\xi}(i,j,k,e,m)=hold1$
      \State $q_{\eta}(i,j,k,e,m)=hold2$
      \State $q_{\zeta}(i,j,k,e,m)=hold3$
    \EndFor
  \EndFor

  \State
  \State $\color{magenta}!\$acc \; end \; data$
  \EndFunction

  \end{algorithmic}
\end{algorithm}

%------------------------------------------------------%
% DSS function
%------------------------------------------------------%

\subsection{Assembly function}

Algorithm \ref{alg:communication} presents pseudocode for the function that performs the direct stiffness summation (DSS) to assemble the global RHS vector. This part of the code requires communication across ranks. The function consists of three major kernels. The first gathering kernels assemble the local RHS vector through overlapping nodes, and therefore requires avoiding race conditions among threads, which is achieved using the \texttt{atomic update} directive. Secondly, the values of the RHS vector at the grid interface are exchanged through MPI communications. The \texttt{host\_data} directive enables the use of device data within the CPU code, MPI calls in this algorithm, by passing the device address of the data to the CPU.

\begin{algorithm}[htb]
  \scriptsize
  \caption{DSS function.}
  \label{alg:communication}
  \begin{algorithmic}
  \Function{PERFORM\underline{\hspace{0.15cm}}DSS}{}

  \State $\color{magenta}!\$acc \; data \; present(rhs_{cg},rhs)$
  \State
  
  \State \textcolor{ForestGreen}{!Kernel: Gathering}
  \State \textcolor{magenta}{$!\$acc \; parallel \; loop \; independent \; collapse(4) \; private(ip\_cg,ip)$}
  \For{$e=1:N_e, k=1:N_{\zeta}, j=1:N_{\eta}, i=1:N_{\xi}$}
    \State $ip\_cg=intma\_cg(i,j,k,e)$
    \State $ip=intma(i,j,k,e)$
    \For{$m=1:N_{\text{var}}$}
      \State \textcolor{magenta}{$!\$acc \; atomic \; update$}      
      \State $rhs_{cg}(m,ip\_cg) = rhs_{cg}(m,ip\_cg) + rhs(m,ip)$
    \EndFor
  \EndFor

  \State

  \State \textcolor{ForestGreen}{!Kernel: Communication}
  \For{$i=1:num\_neighbors$}
    \State $Pack \; send\_data$
    \State \textcolor{magenta}{$!\$acc \; host\_data \; use\_device(recv\_data) $}
    \State $MPI\_Irecv(recv\_data,num\_send\_recv,i\_dest)$
    \State \textcolor{magenta}{$!\$acc \; end \; host\_data $}
    \State
    \State \textcolor{magenta}{$!\$acc \; host\_data \; use\_device(send\_data) $}
    \State $MPI\_Isend(send\_data,num\_send\_recv,i\_dest)$
    \State \textcolor{magenta}{$!\$acc \; end \; host\_data $}
    \State $MPI\_Waitall()$
    \State $Add \; recv\_data \; to\; rhs_{cg}$
  \EndFor

  \State

  \State \textcolor{ForestGreen}{!Kernel: Scattering}
  \State \textcolor{magenta}{$!\$acc \; parallel \; loop \; independent \; collapse(4) \; private(ip\_cg,ip)$}
  \For{$e=1:N_e, k=1:N_{\zeta}, j=1:N_{\eta}, i=1:N_{\xi}$}
    \State $ip\_cg=intma\_cg(i,j,k,e)$
    \State $ip = intma(i,j,k,e)$
    \For{$m=1:N_{\text{var}}$}
      \State $rhs(m,ip) = rhs_{cg}(m,ip\_cg)$
    \EndFor
  \EndFor

  \State
  \State $\color{magenta}!\$acc \; end \; data$
  \EndFunction

  \end{algorithmic}
\end{algorithm}

%------------------------------------------------------%
%   								Results
%------------------------------------------------------%

\section{Simulation setup}\label{sec:setup}

\subsection{Numerical setup}

We simulate a tropical cyclone case using the GPU-accelerated code. The computational domain is a 3D box with horizontal dimensions of 800 km and a vertical height of 20 km, defined as $\Omega=[-400,400]\times[-400,400]\times[0,20]$ km. Doubly periodic boundary conditions are applied at the lateral boundaries, and a no-flux boundary condition is enforced at the bottom surface. To absorb upward gravity waves and prevent reflections, an implicit sponge layer \cite{klemp2008upper, kang2024multiscale} with a thickness of 4 km is imposed at the top of the domain. 

Several values for horizontal grid spacing are tested, with $\dx\in\qty{0.5, 1, 2, 4}$ km when using fifth-order basis functions $(N=5)$, while the vertical spacing is fixed at $\dz=167$ m. For different orders of the basis function, the grid spacing is adjusted as close as possible to these values. The simulation runs for 6 hours with a time step size of $\dt=0.5$ seconds.

% Sounding -- theta, pressure

\begin{figure}[h!]
	\captionsetup[subfigure]{justification=centering}
  \centering
  \begin{subfigure}[b]{0.24\textwidth}
    \includegraphics[width=\textwidth]{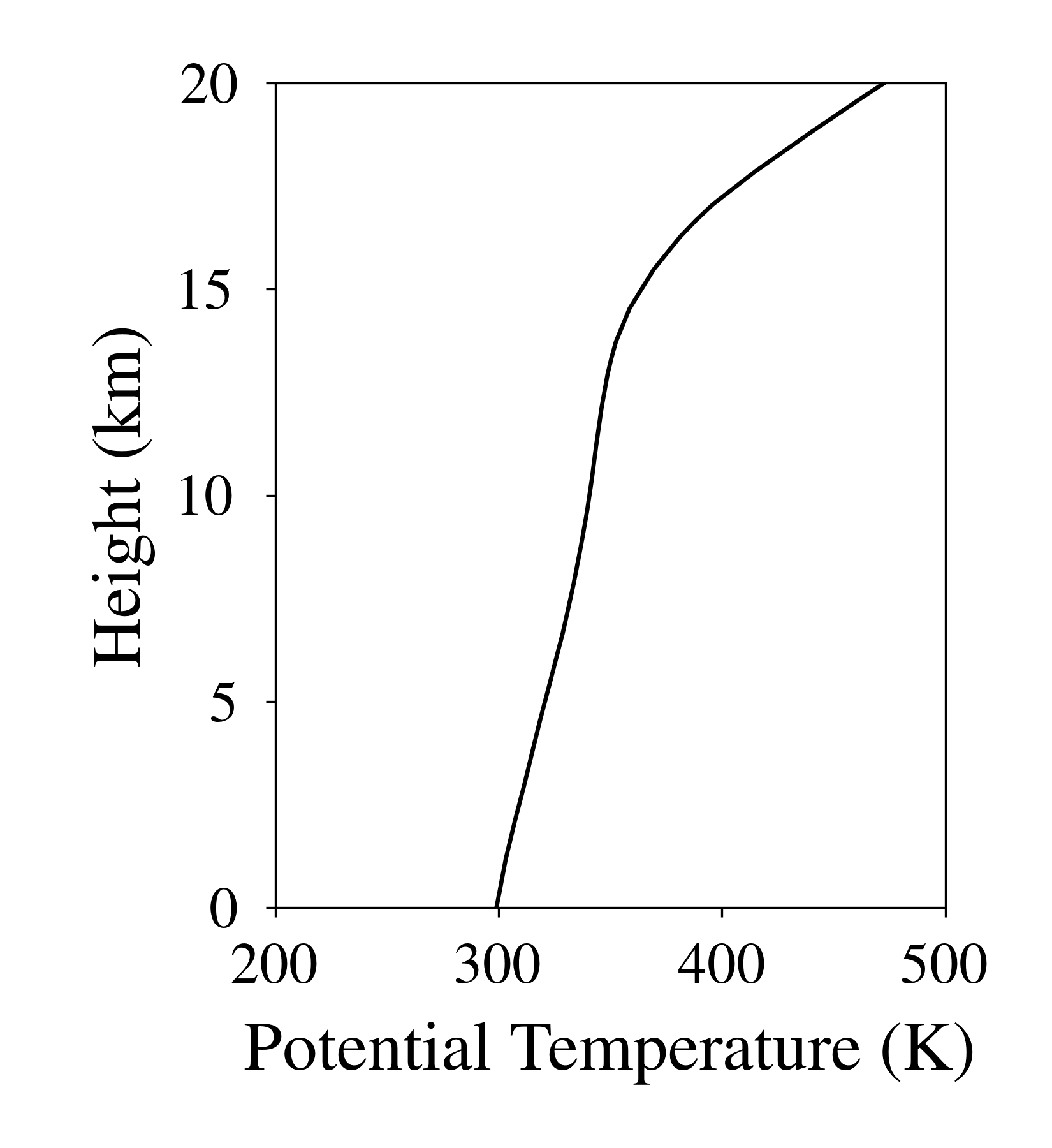}
    \caption{$\theta_0$}
  \end{subfigure}
  \begin{subfigure}[b]{0.24\textwidth}
    \includegraphics[width=\textwidth]{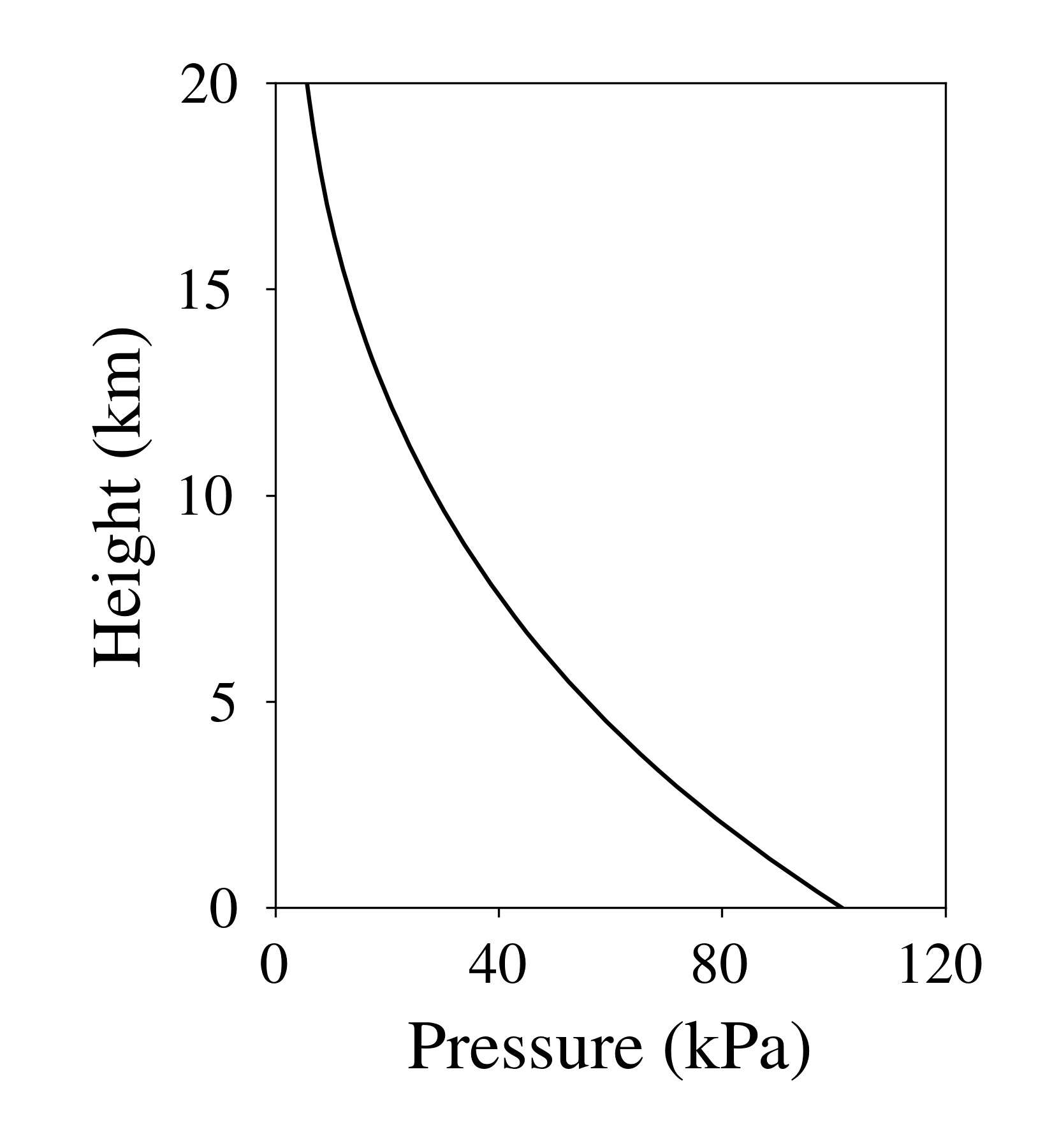}
    \caption{$p_0$}
  \end{subfigure} 
  \caption{Soundings for tropical cyclone simulations: reference potential temperature $\theta_0$ and reference pressure $p_0$.}\label{fig:sounding}
\end{figure}

The values for the reference variables are taken from the background sounding provided in \cite{jordan1958mean}. Figure \ref{fig:sounding} presents the profiles of reference potential temperature and reference pressure with respect to height. In the simulation, background wind is not considered.

The tropical cyclone is initiated by an idealized vortex used in \citep{guimond2016impacts}. This vortex is modified from the original form \citep{nolan2003nonhydrostatic,nolan2007tropical} to enforce zero velocities at the lateral boundaries. In this vortex, the azimuthal-mean tangential velocity is defined as
\begin{equation}\label{eq:initial-velocity}
  \overline{u}_{\theta}(r,z) = V(r) \exp\qty(-\dfrac{z^{\sigma}}{\sigma D^{\sigma}-1}) \exp\qty[-\qty(\dfrac{r}{D_2})^6],
\end{equation}
where $V$ is the surface tangential velocity, $\sigma=2$, $D_1=5823$ m, and $D_2=200$ km. The surface tangential velocity is calculated by integrating a Gaussian distribution of vorticity with a peak of $1.5\times 10^{-3}$ s$^{-1}$ at the vortex center, and a maximum wind speed of $21.5$ m/s at a radius of 50 km.

\begin{figure}[h!]
	\captionsetup[subfigure]{justification=centering}
  \centering
  \begin{subfigure}[b]{0.24\textwidth}
    \includegraphics[width=\textwidth]{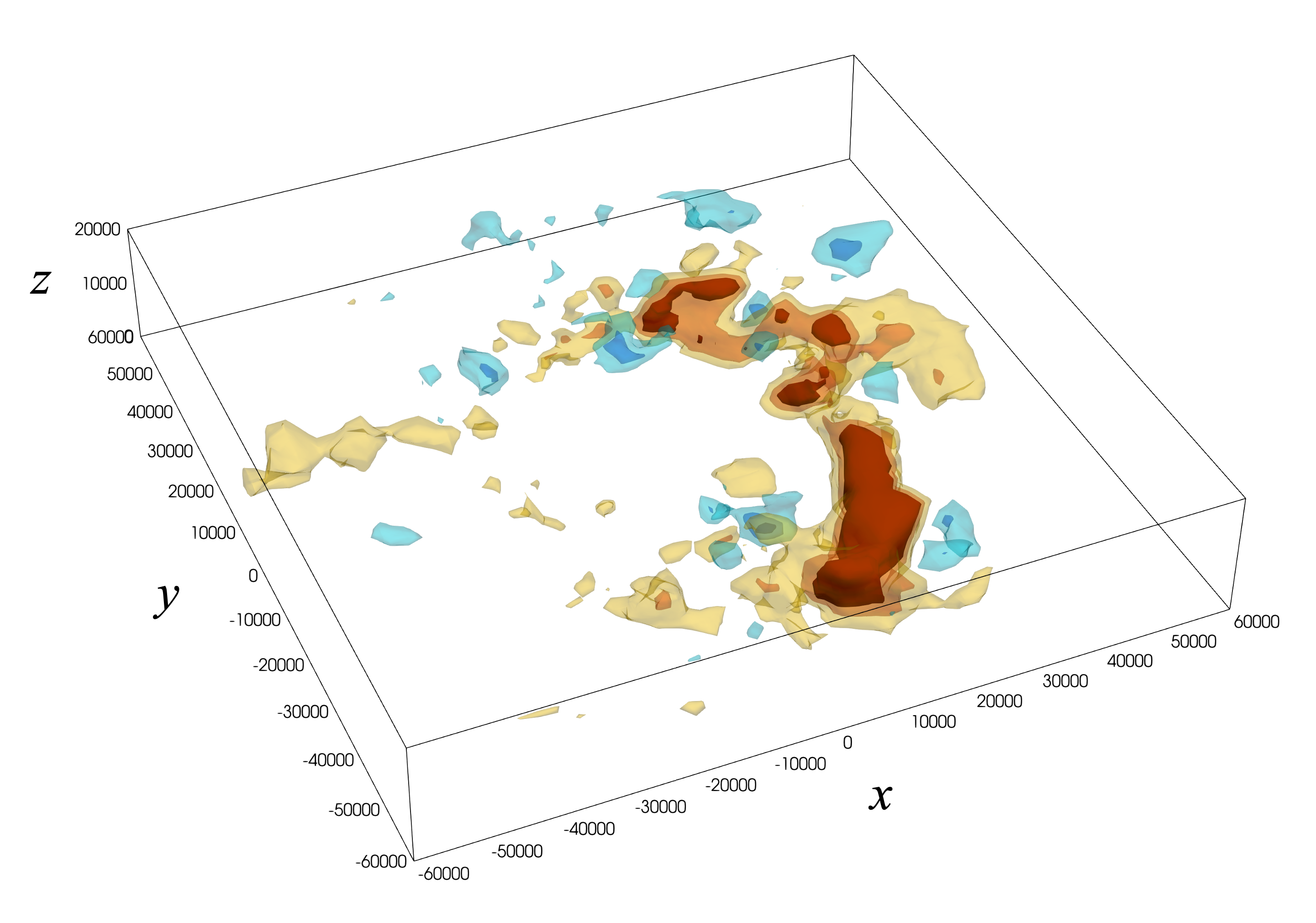}
    \caption{$t=0.5$ hours}
  \end{subfigure}
  \begin{subfigure}[b]{0.24\textwidth}
    \includegraphics[width=\textwidth]{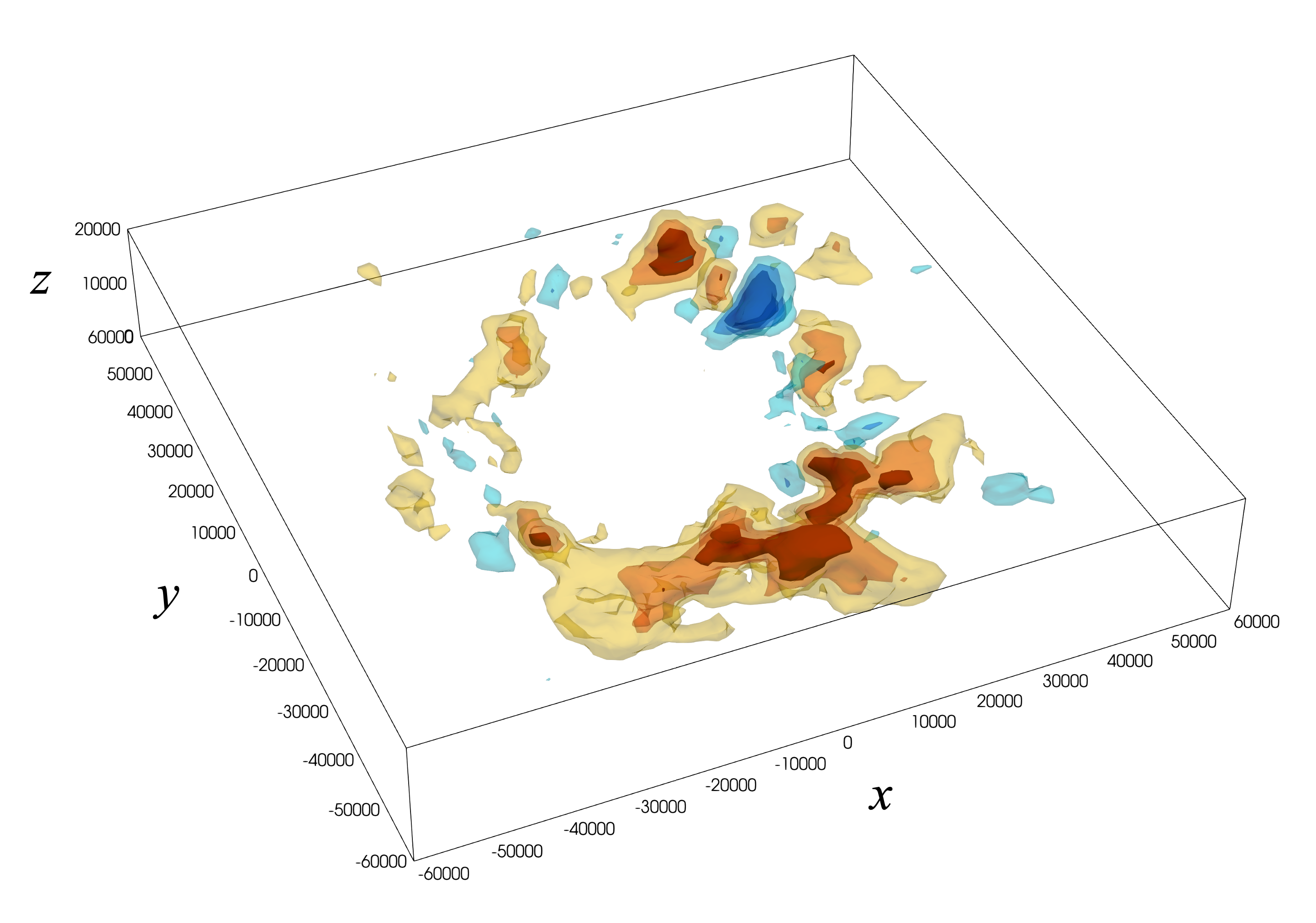}
    \caption{$t=1.633$ hours}
  \end{subfigure} 
  \begin{subfigure}[b]{0.24\textwidth}
    \includegraphics[width=\textwidth]{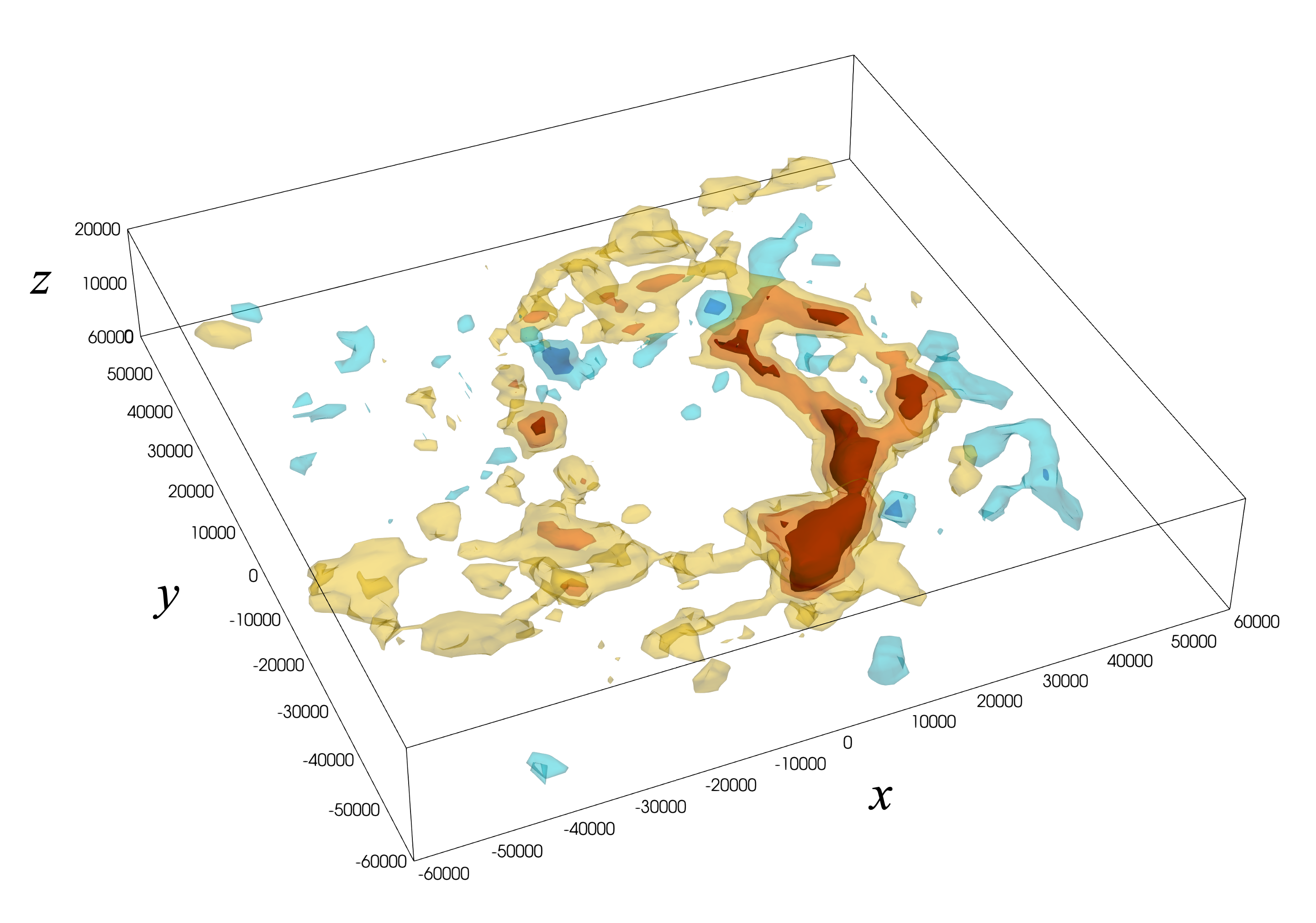}
    \caption{$t=2.767$ hours}
  \end{subfigure} 
  \begin{subfigure}[b]{0.24\textwidth}
    \includegraphics[width=\textwidth]{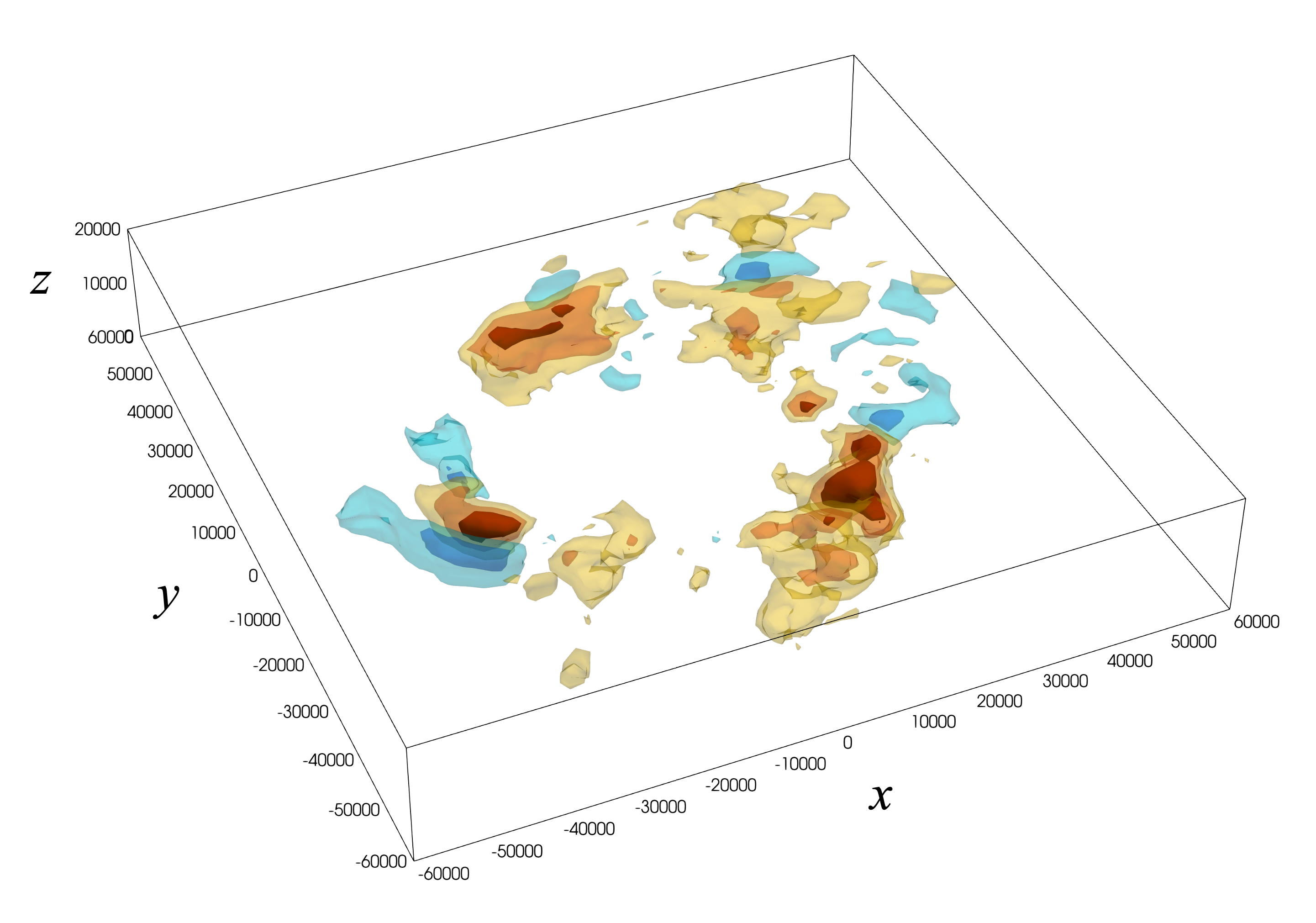}
    \caption{$t=3.9$ hours}
  \end{subfigure} 
  \begin{subfigure}[b]{0.24\textwidth}
    \includegraphics[width=\textwidth]{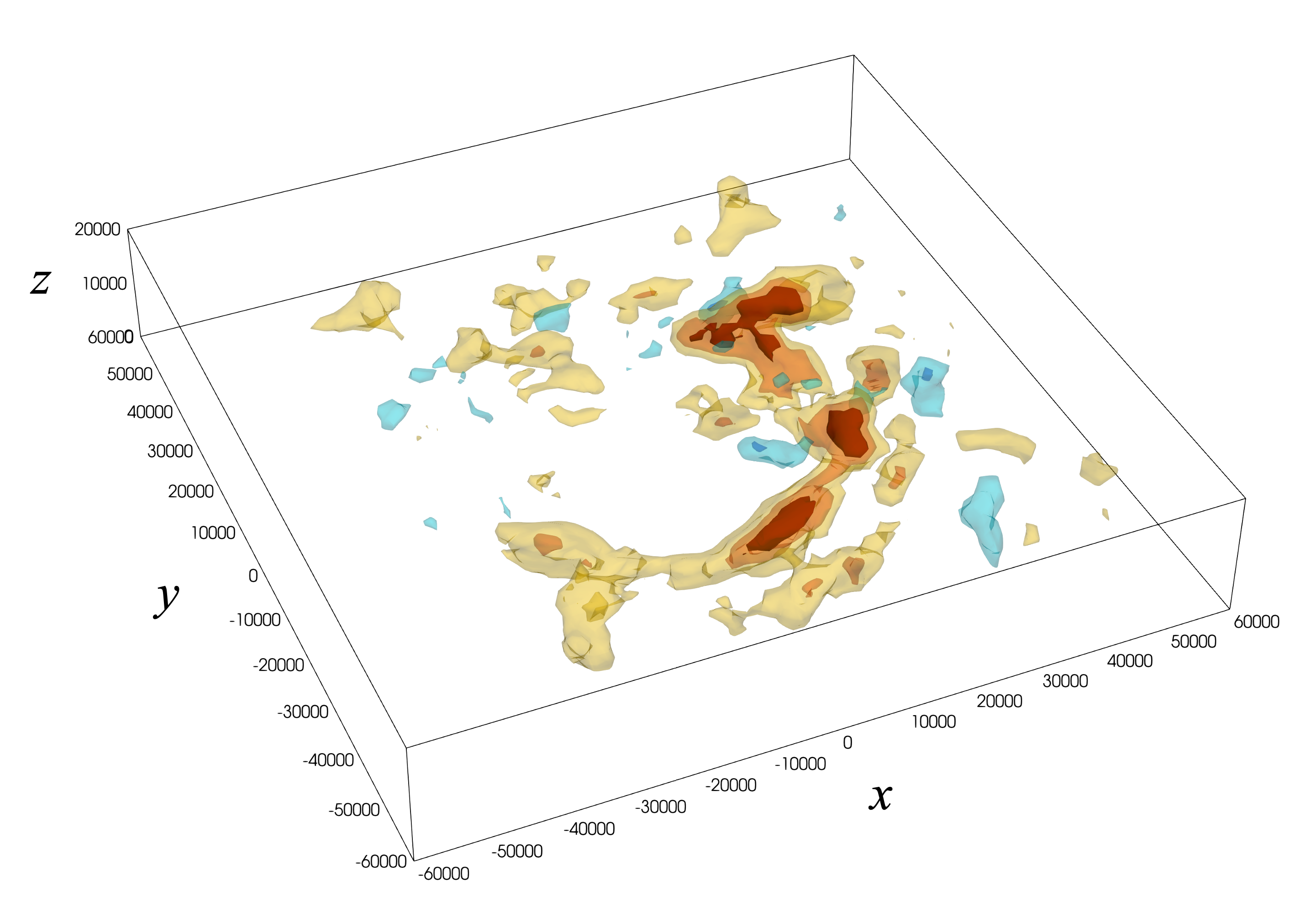}
    \caption{$t=5.033$ hours}
  \end{subfigure} 
  \begin{subfigure}[b]{0.24\textwidth}
    \includegraphics[width=\textwidth]{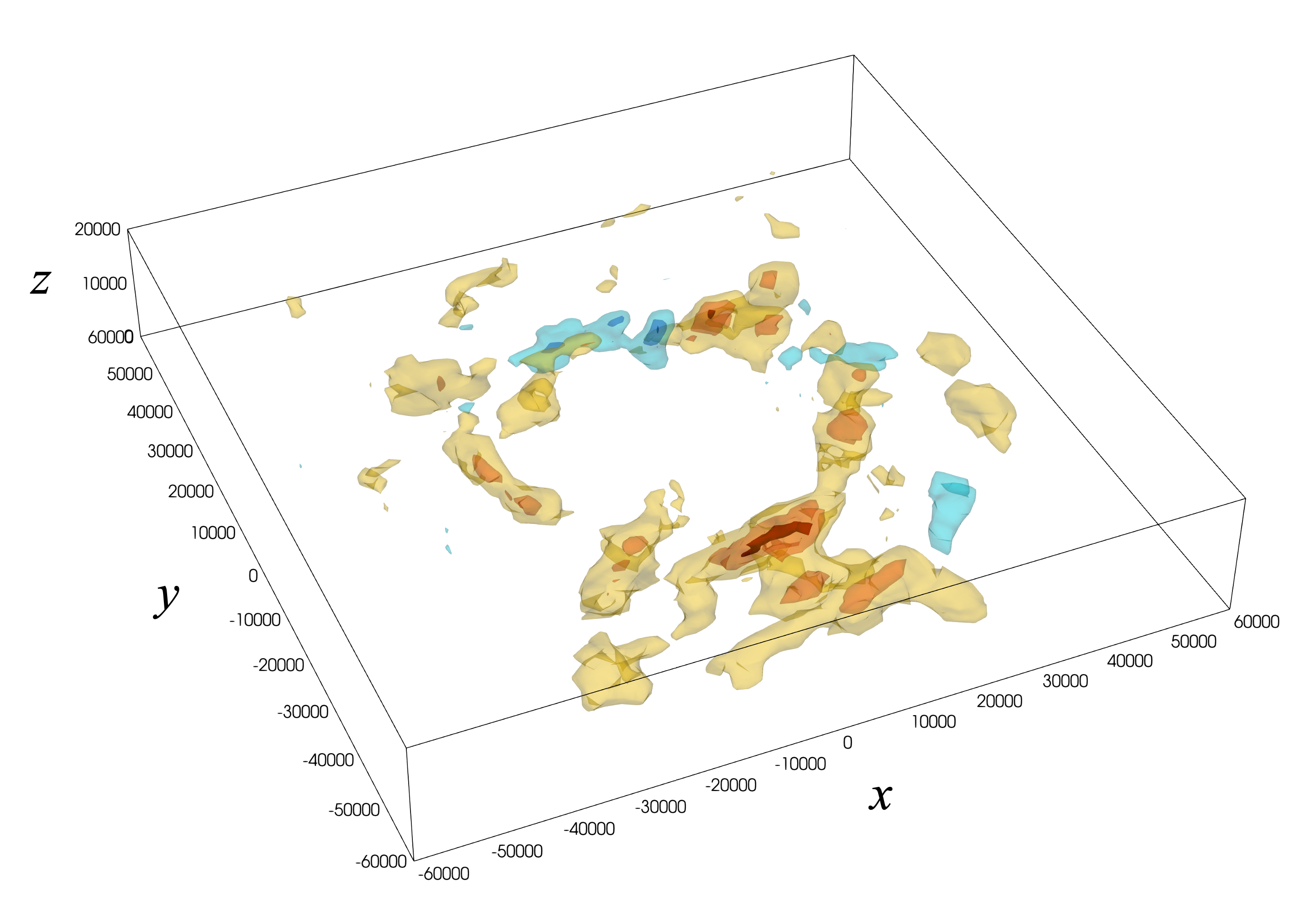}
    \caption{$t=5.6$ hours}
  \end{subfigure}
  \\
  \vspace{0.5em}
  \begin{subfigure}[b]{0.24\textwidth}
    \includegraphics[width=\textwidth]{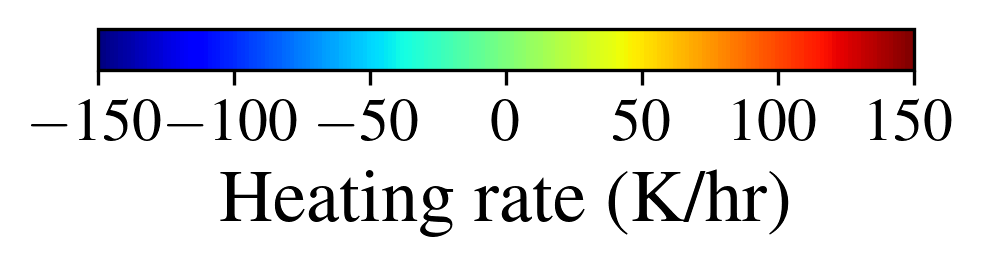}
  \end{subfigure}
  \caption{Latent heating in the region $[-60,60]\times[-60,60]\times[0,20]$ km.}\label{fig:heating}
\end{figure}

The RI of tropical cylones is driven by the vortex response to latent heating injected into the boundary layer. In this research, we incorporate the observational latent heating into the source term $S_{\theta}$ in Eq.\ \ref{eq:governing:theta}. A time-varying 3D observational heating profile was obtained from airborne Doppler radar measurements during the RI of Hurricane Guillermo (1997) \citep{hasan2022effects}. Figure \ref{fig:heating} presents snapshots of the asymmetric spatial distribution of latent heating at different time points. Over the duration of 6 hours, 11 sampling points of latent heating are used to interpolate the heating source term.

\subsection{Computing system}

The performance of the GPU code for tropical cyclone simulations is evaluated on the Delta system, which was ranked No.\ 265 in the November 2024 TOP 500 list \citep{top500_2024}. Delta is an advanced computing resource maintained by the National Center for Supercomputing Applications (NCSA) at the University of Illinois, and supported by the National Science Foundation (NSF). While Delta provides access to both NVIDIA A40 GPUs and A100 GPUs, we only use A100 GPUs for performance tests in this study. Delta features 100 A100 GPU nodes, and each node is equipped with 64 AMD EPYC 7763 (Milan) CPUs and 4 A100 GPUs. The compute nodes are connected via a low-latency high-bandwidth HPE/Cray Slingshot interconnect.

% , and GPUs are interconnected through NVLinks. 

The parallel code is compiled using NVIDIA HPC compilers (version 22.5), which include OpenMPI compiler wrappers. All performance tests in this study are conducted using double-precision calculations.

\section{Results}\label{sec:results}

\subsection{Numerical results}

\begin{figure}[h!]
	\captionsetup[subfigure]{justification=centering}
  \centering
  \begin{subfigure}[b]{0.48\textwidth}
    \includegraphics[width=\textwidth]{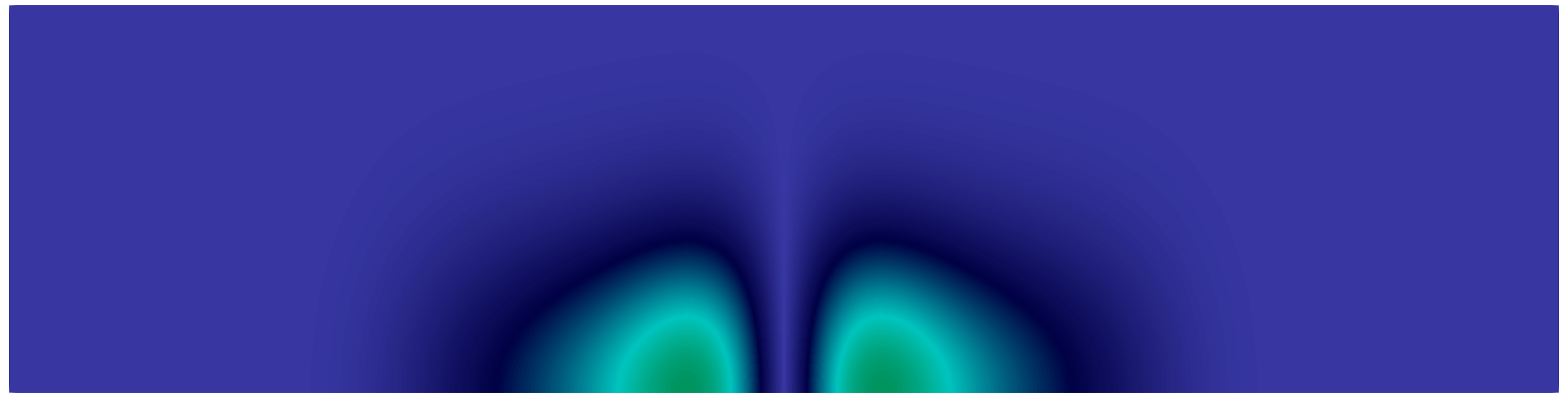}
    \caption{$t=0$ hours}
  \end{subfigure}
  \begin{subfigure}[b]{0.48\textwidth}
    \includegraphics[width=\textwidth]{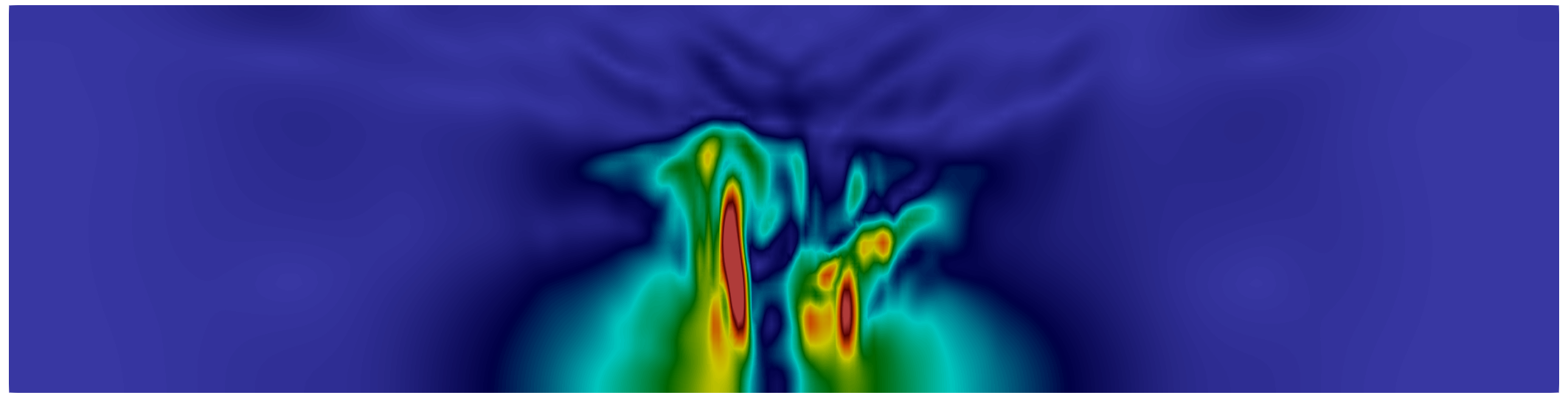}
    \caption{$t=2$ hours}
  \end{subfigure}
  \begin{subfigure}[b]{0.48\textwidth}
    \includegraphics[width=\textwidth]{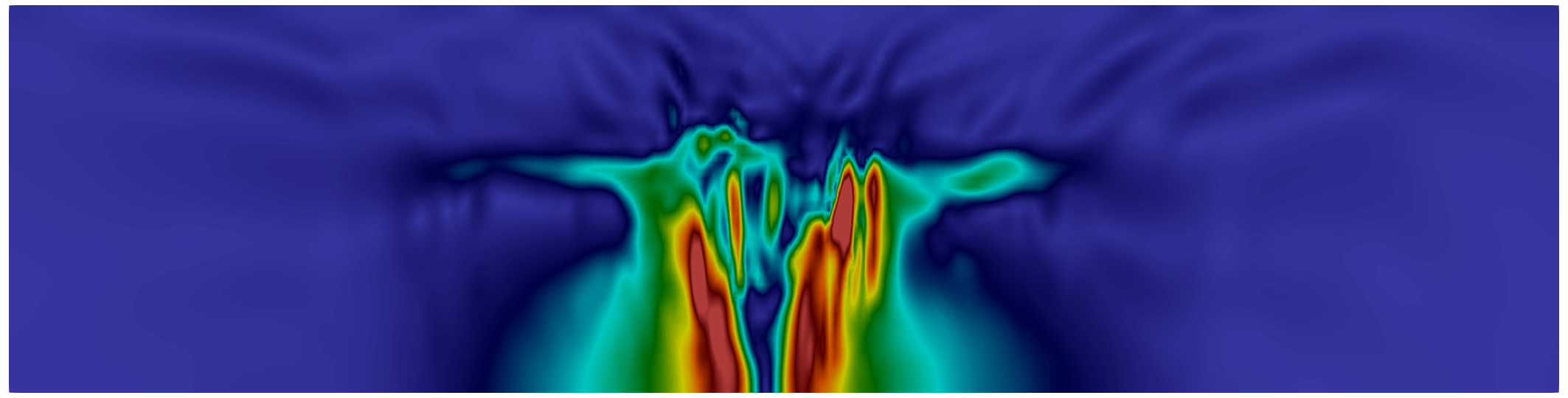}
    \caption{$t=4$ hours}
  \end{subfigure}
  \begin{subfigure}[b]{0.48\textwidth}
    \includegraphics[width=\textwidth]{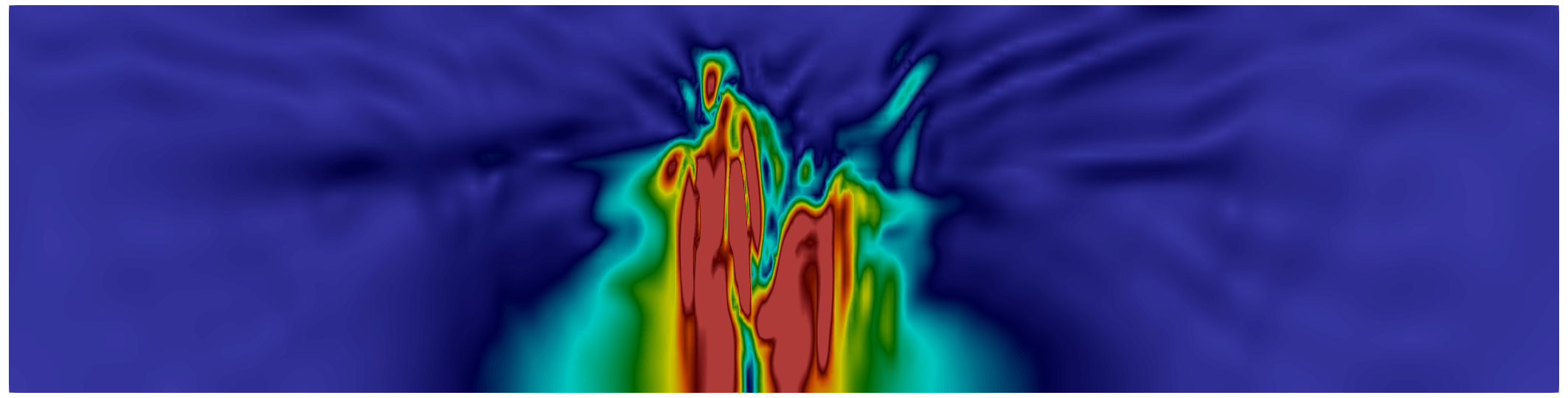}
    \caption{$t=6$ hours}
  \end{subfigure}  
  \\
  \begin{subfigure}[b]{0.18\textwidth}
    \includegraphics[width=\textwidth]{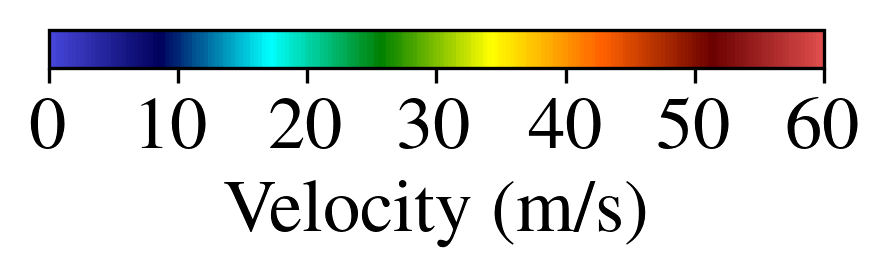}
  \end{subfigure} 
  \caption{Velocity magnitude at the middle $y$-$z$ plane. (The domain is vertically stretched by a factor of 10 for visual purposes.)}\label{fig:instance-vel}
\end{figure}

\begin{figure}[h!]
	\captionsetup[subfigure]{justification=centering}
  \centering
  \begin{subfigure}[b]{0.48\textwidth}
    \includegraphics[width=\textwidth]{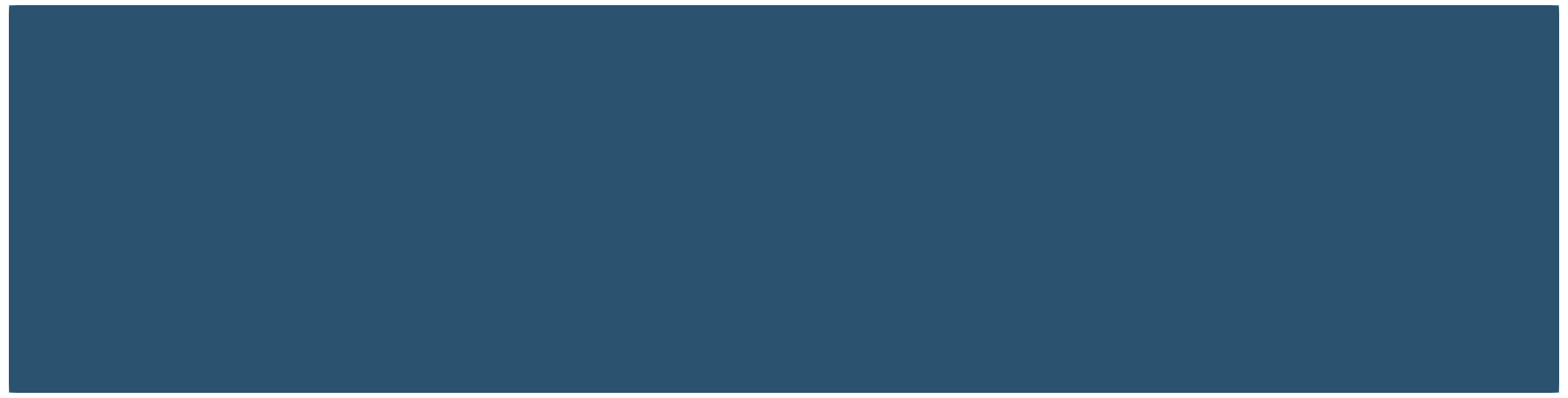}
    \caption{$t=0$ hours}
  \end{subfigure}
  \begin{subfigure}[b]{0.48\textwidth}
    \includegraphics[width=\textwidth]{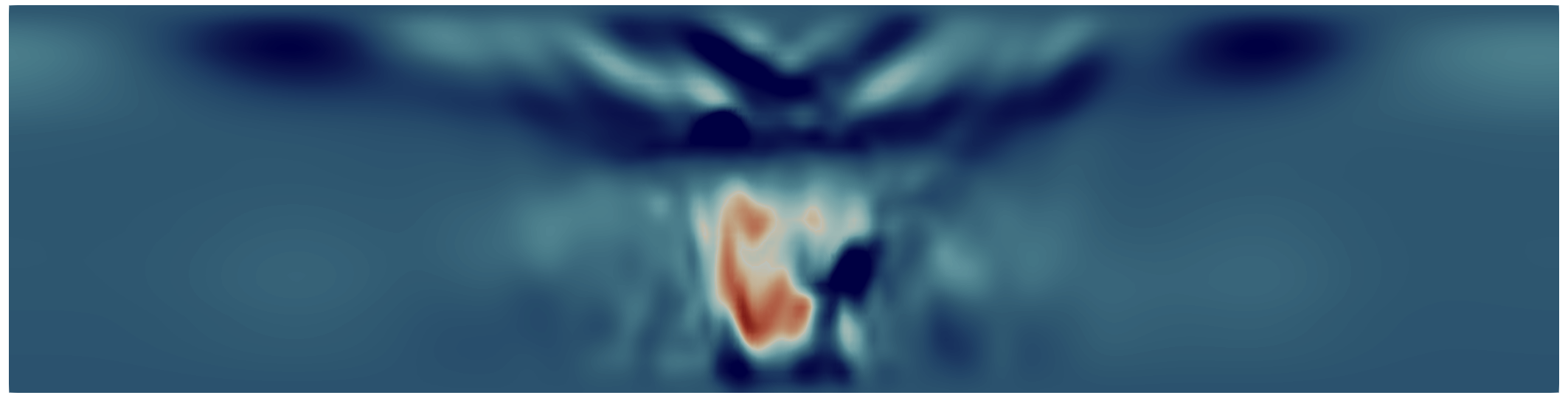}
    \caption{$t=2$ hours}
  \end{subfigure}
  \begin{subfigure}[b]{0.48\textwidth}
    \includegraphics[width=\textwidth]{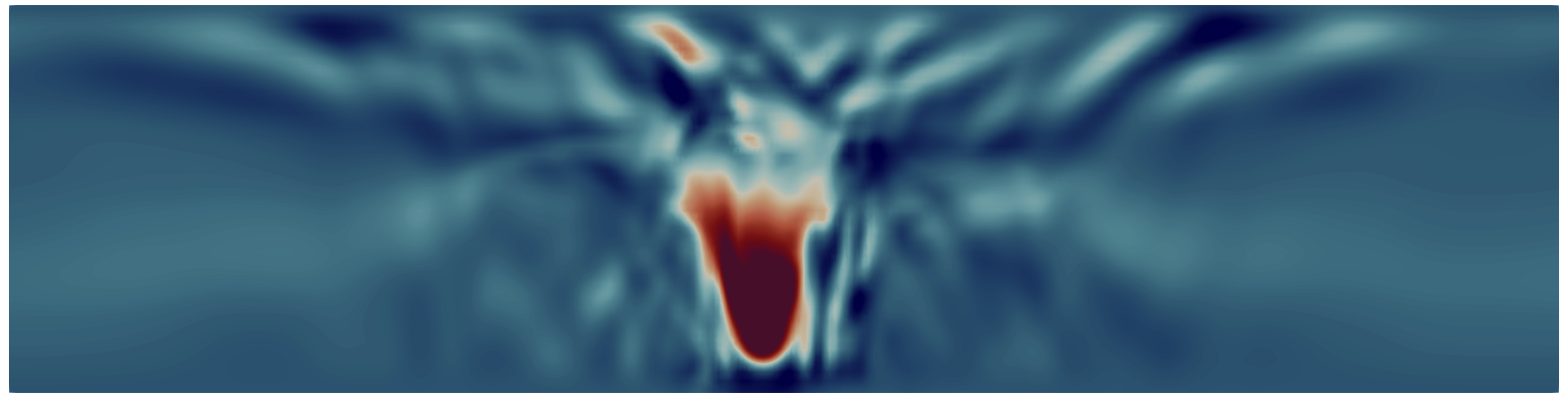}
    \caption{$t=4$ hours}
  \end{subfigure}
  \begin{subfigure}[b]{0.48\textwidth}
    \includegraphics[width=\textwidth]{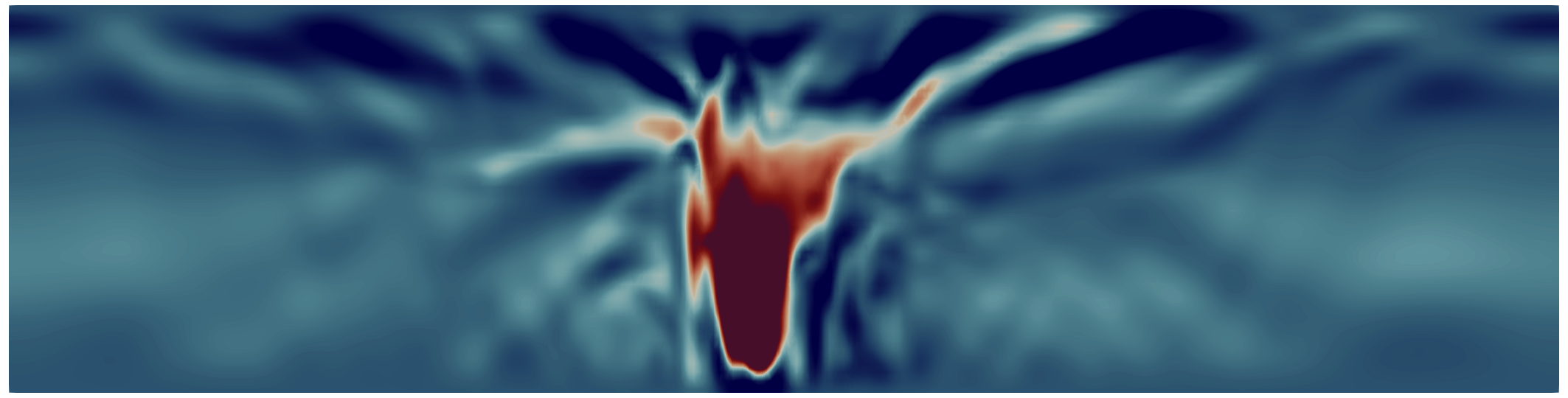}
    \caption{$t=6$ hours}
  \end{subfigure}  
  \\
  \begin{subfigure}[b]{0.18\textwidth}
    \includegraphics[width=\textwidth]{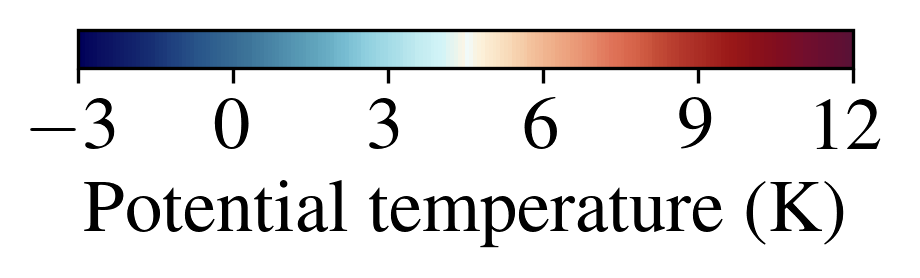}
  \end{subfigure} 
  \caption{Potential temperature perturbation at the middle $y$-$z$ plane. (The domain is vertically stretched by a factor of 10 for visual purposes.)}\label{fig:instance-theta}
\end{figure}

\begin{figure}[h!]
	\captionsetup[subfigure]{justification=centering}
  \centering
  \begin{subfigure}[b]{0.44\textwidth}
    \includegraphics[width=\textwidth]{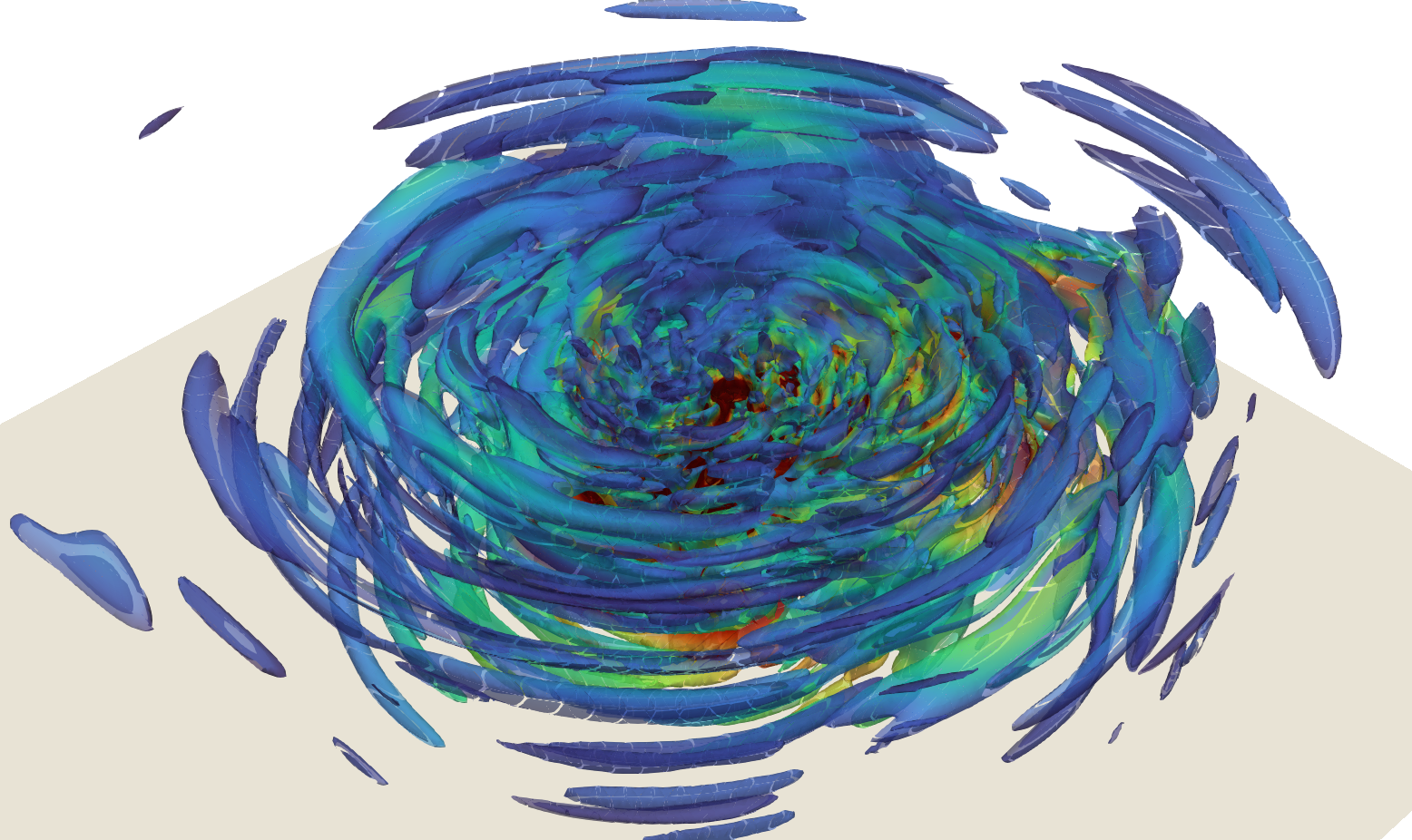}
  \end{subfigure}
  \\
  \begin{subfigure}[b]{0.18\textwidth}
    \includegraphics[width=\textwidth]{figures/instantaneous/Figure_Colormap_Velocity.png}
  \end{subfigure} 
  \caption{Vortical structures in the eye and spiral bands of the tropical cyclone at $t=6$ hours. (Vortices are visualized using the $Q$-criterion.)}\label{fig:Qcriterion}
\end{figure}

Figures \ref{fig:instance-vel} and \ref{fig:instance-theta} show the velocity magnitude and potential temperature perturbation, respectively, on the $y$-$z$ plane at the center of the domain. In the initial state at $t=0$ hours, a single idealized vortex defined by Eq.\ \ref{eq:initial-velocity} is located at the center near the surface, with no thermal perturbation. By injecting thermal energy into the dynamical system, a strong updraft develops in the eyewall and leads to asymmetric and complex vortex structures. Figure \ref{fig:Qcriterion} shows the vortical structures in the eye and spiral bands of the tropical cyclone at $t=6$ hours.

% Instantaneous velocity

% Instantaneous theta

% Instantaneous Q-isosurface

\begin{figure}[h!]
	\captionsetup[subfigure]{justification=centering}
  \centering
  \begin{subfigure}[b]{0.24\textwidth}
    \includegraphics[width=\textwidth]{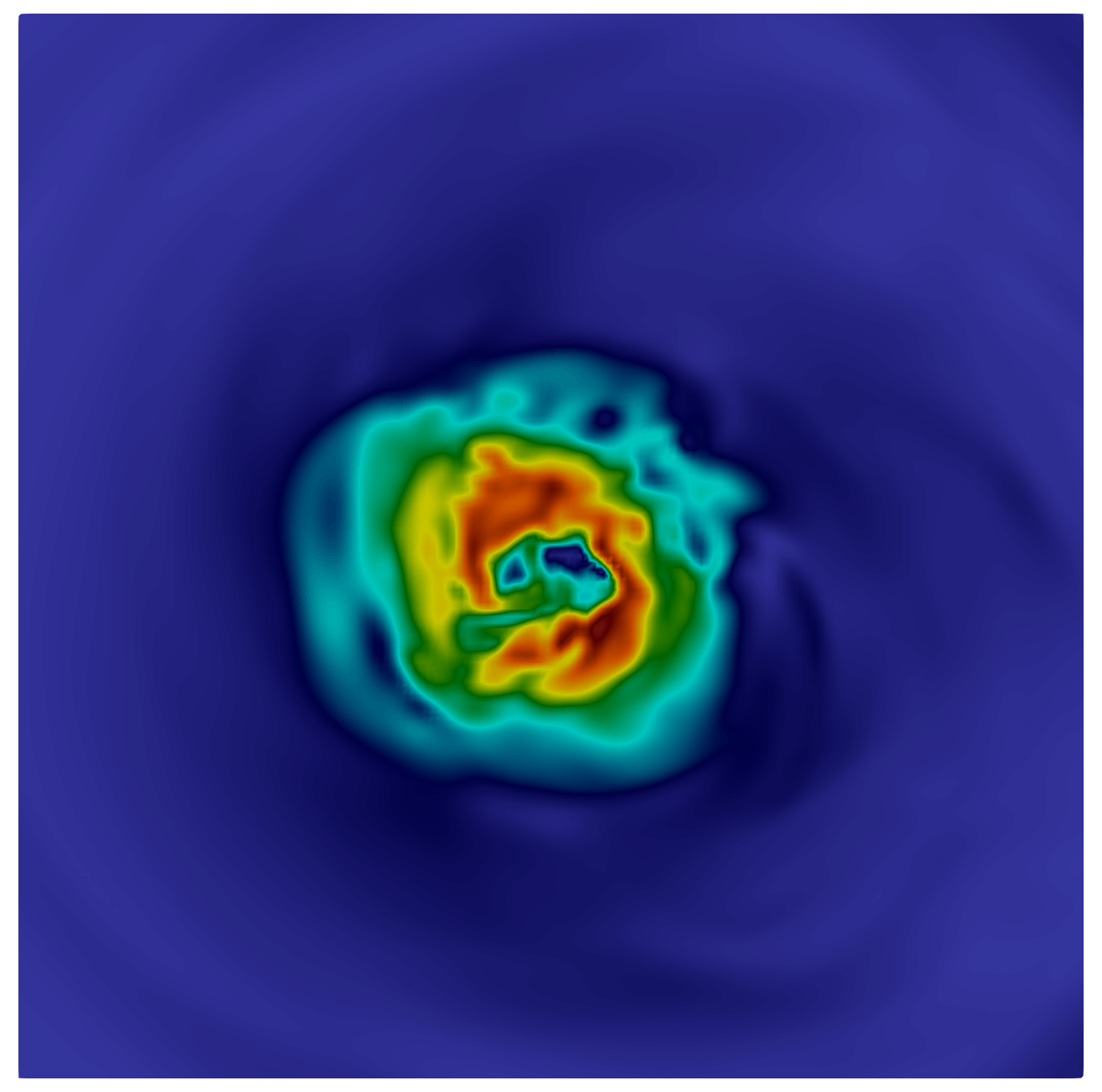}
    \caption{$z=8$ km}
  \end{subfigure}
  \begin{subfigure}[b]{0.24\textwidth}
    \includegraphics[width=\textwidth]{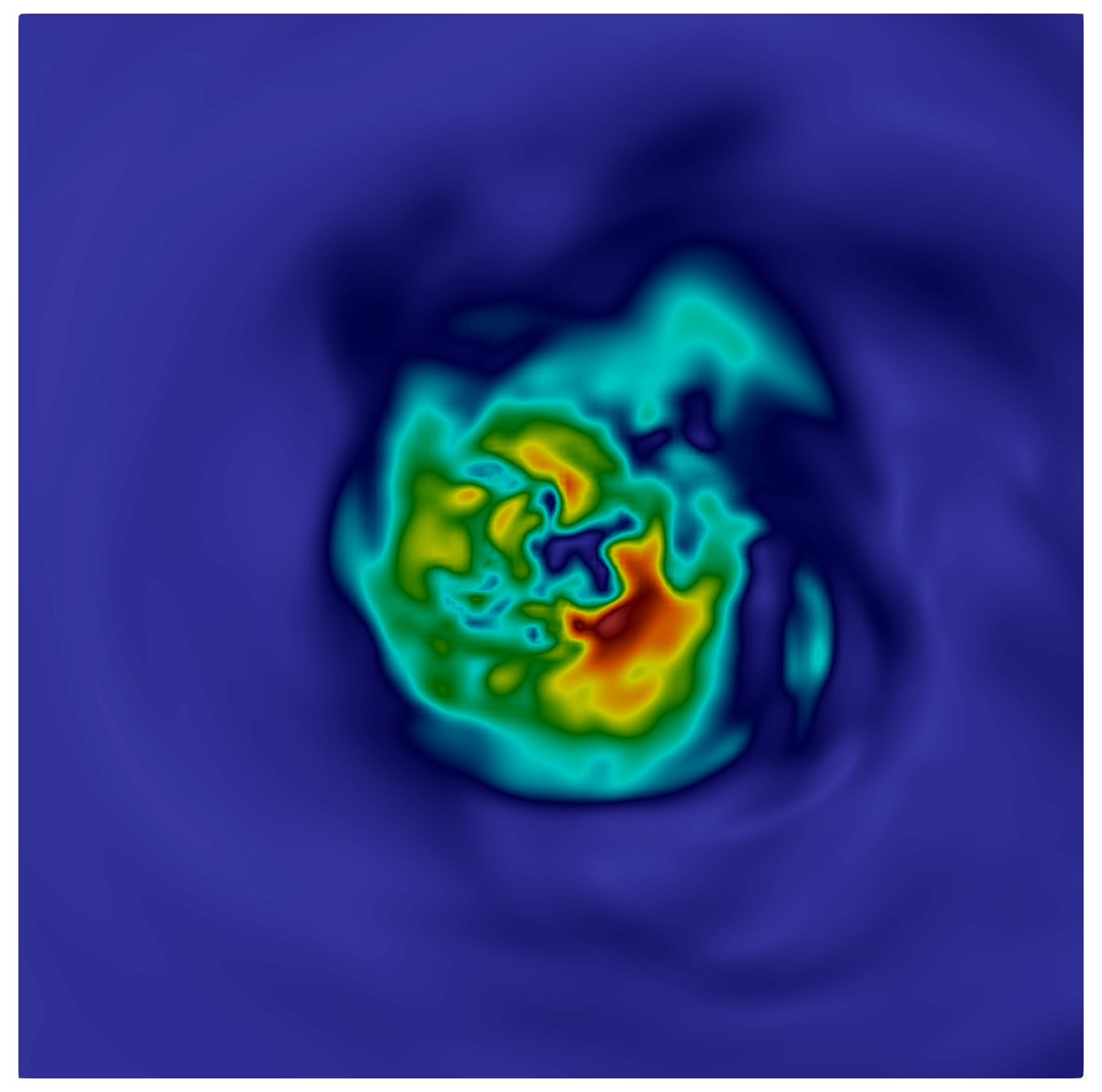}
    \caption{$z=10$ km}
  \end{subfigure}
  \begin{subfigure}[b]{0.24\textwidth}
    \includegraphics[width=\textwidth]{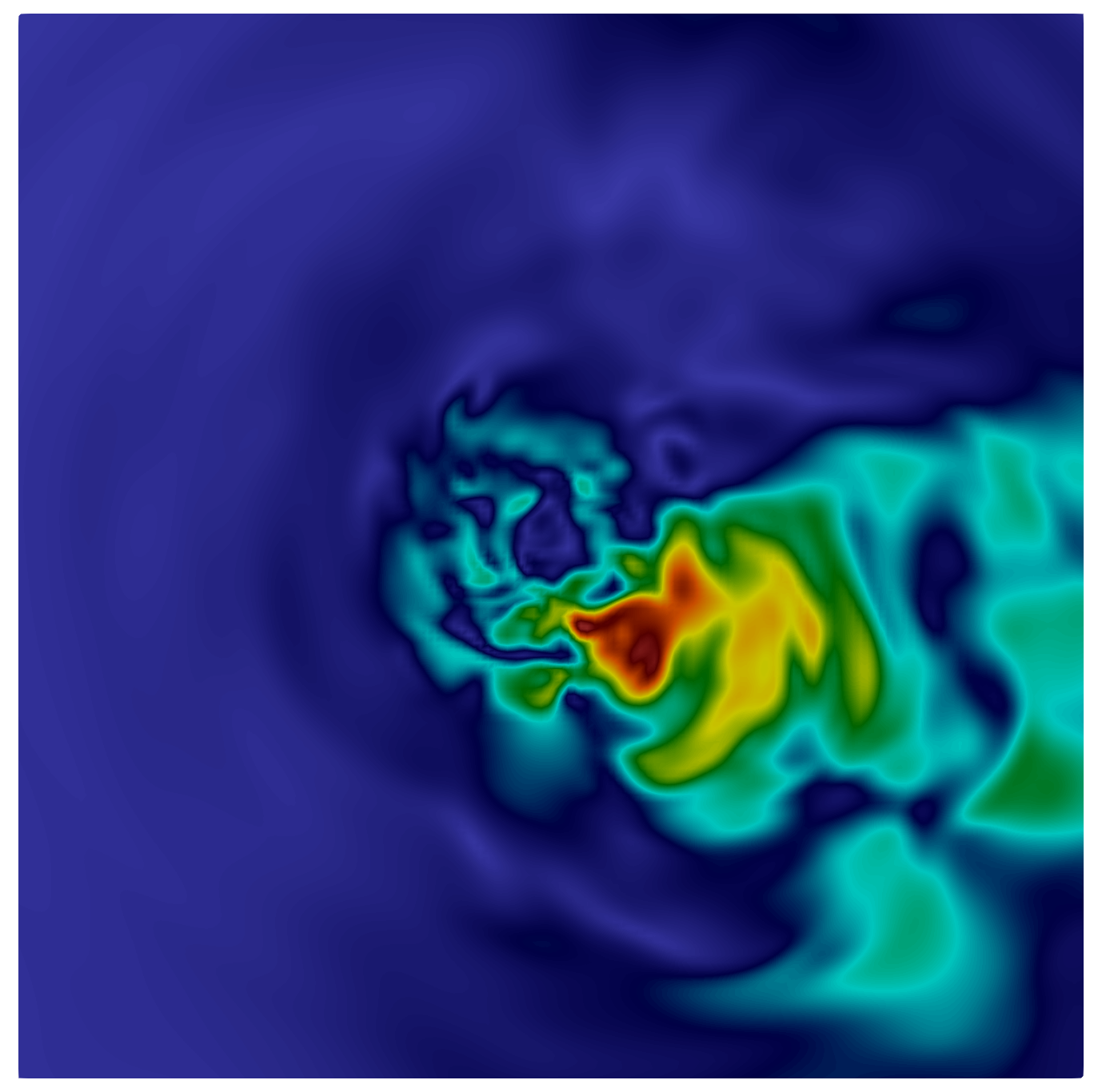}
    \caption{$z=12$ km}
  \end{subfigure}
  \begin{subfigure}[b]{0.24\textwidth}
    \includegraphics[width=\textwidth]{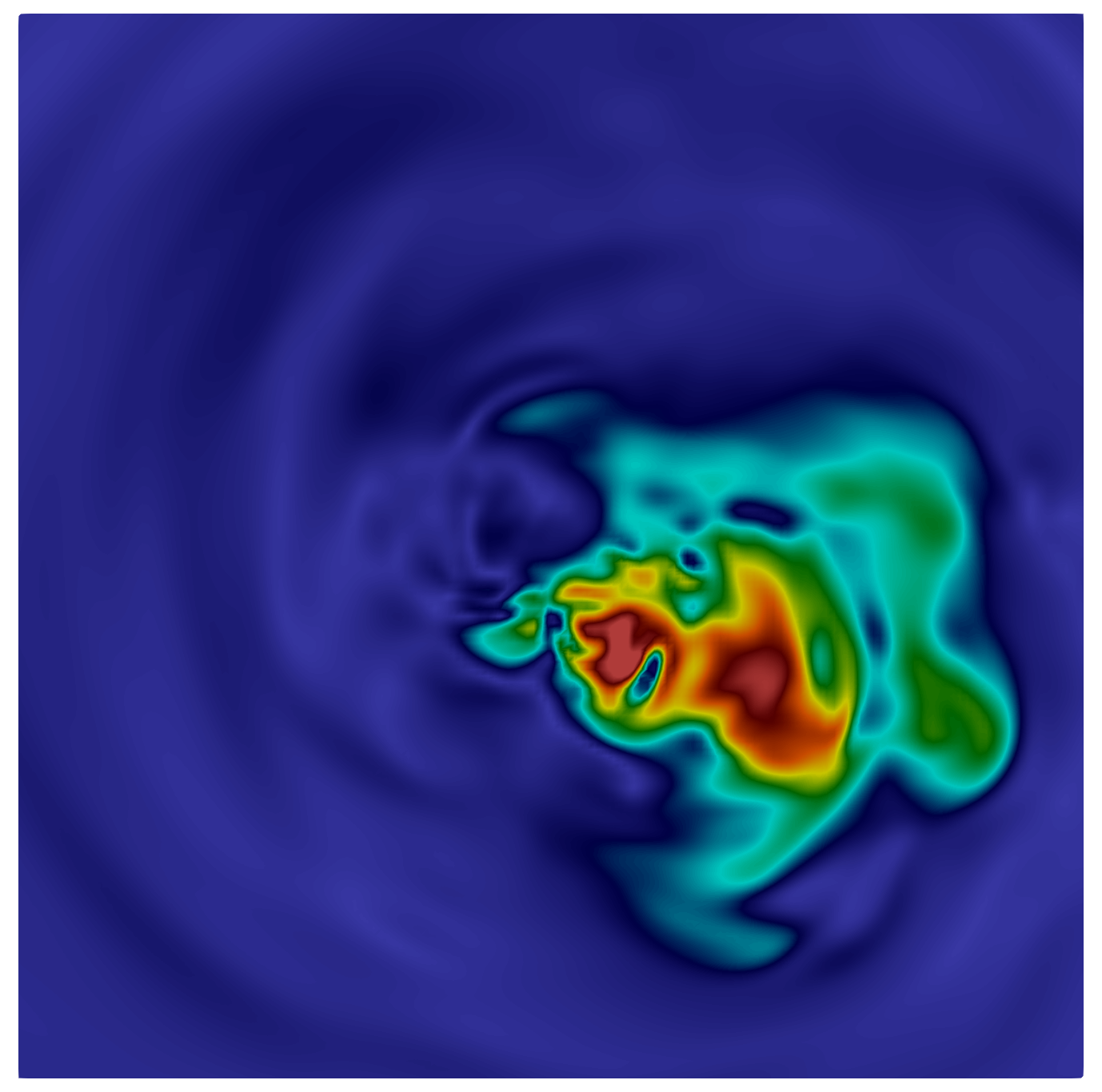}
    \caption{$z=14$ km}
  \end{subfigure}  
  \\
  \begin{subfigure}[b]{0.18\textwidth}
    \includegraphics[width=\textwidth]{figures/instantaneous/Figure_Colormap_Velocity.png}
  \end{subfigure} 
  \caption{Velocity magnitude at $t=6$ hours in the horizontal region $[-20,20]\times[-20,20]$ km at various heights.}\label{fig:vel-xy}
\end{figure}

Figure \ref{fig:vel-xy} shows the velocity magnitude on the horizontal plane at various heights. Uneven temperature distribution near the surface significantly contributes to the asymmetric structure of the tropical cyclone, where the extent of asymmetry varies with height.

% Instantaneous velocity at X-Y plane

%------------------------------------------------------%
%   								RI velocity
%------------------------------------------------------%

In order to quantify the magnitude of the vortex, we use the index of maximum tangential velocity and denote it as the RI velocity. In this work, the RI velocity is defined as the maximum of azimuthal-averaged tangential velocity as follows:
\begin{equation}\label{eq:u_RI}
  u_{\text{RI}} = \max\qty\big(\overline{u}_{\theta}(r,z)),
\end{equation}
where
\begin{align}
  \overline{u}_{\theta}(r,z) &= \frac{1}{2\pi}\int^{2\pi}_{0} u_{\theta} (r,\theta,z)d\theta \label{eq:azimuthal-average}\\
  u_{\theta} &= \norm{\ub_{\theta}} \\
  \ub_{\theta} &= \qty(\Ib-\eb_r\otimes\eb_r - \eb_z\otimes\eb_z)\ub.
\end{align}
Here, $\eb_r$ and $\eb_z$ are the unit vectors in the radial and vertical directions, respectively, which are calculated as $\eb_r=\rb/\norm{\rb}$, $\rb=(x-x_c,y-y_c,0)$, and $\eb_z=(0,0,1)$. The vertically varying centroid of the TC, $\xb_c=(x_c,y_c,z)$, is estimated as the weighted center of the velocity magnitude field:
\begin{equation}
  \xb_c = \frac{\int_x \int_y \; \norm{\ub} \xb \; dx dy}{\int_x\int_y \norm{\ub} \; dx dy},
\end{equation}
where $L_x$ and $L_y$ are the lengths of the domain along the $x$ and $y$ directions, respectively. 

We propose a simplified calculation of azimuthally averaged velocity, as given in Eq.\ \ref{eq:azimuthal-average}. The azimuthally averaged velocity is approximated as the arithmetic mean of points that fall within a narrow circular strip at each radial distance $r_1$ and height $z_1$. This circular strip is defined as $\Omega_{\text{strip}}(r_1,z_1)=\qty\big{(r,\theta,z) \mid r_1-\tfrac{\varepsilon}{2} < r < r_1+\tfrac{\varepsilon}{2}, 0\le\theta\le 2\pi, z=z_1}$, where $\varepsilon$ is the width of the strip. The averaged tangential velocity is evaluated as
\begin{equation}
  \overline{u}_{\theta}(r,z) \approx \frac{1}{N_p}\sum_{i=1}^{N_p} u_{\theta}(r_i,\theta_i,z_i), \;\text{for}\; (r_i,\theta_i,z_i) \in \Omega_{\text{strip}}(r,z),
\end{equation}
where $N_p$ is the number of points that belong to the strip.

\begin{figure}[h!]
  \centering
  \includegraphics[width=0.45\textwidth]{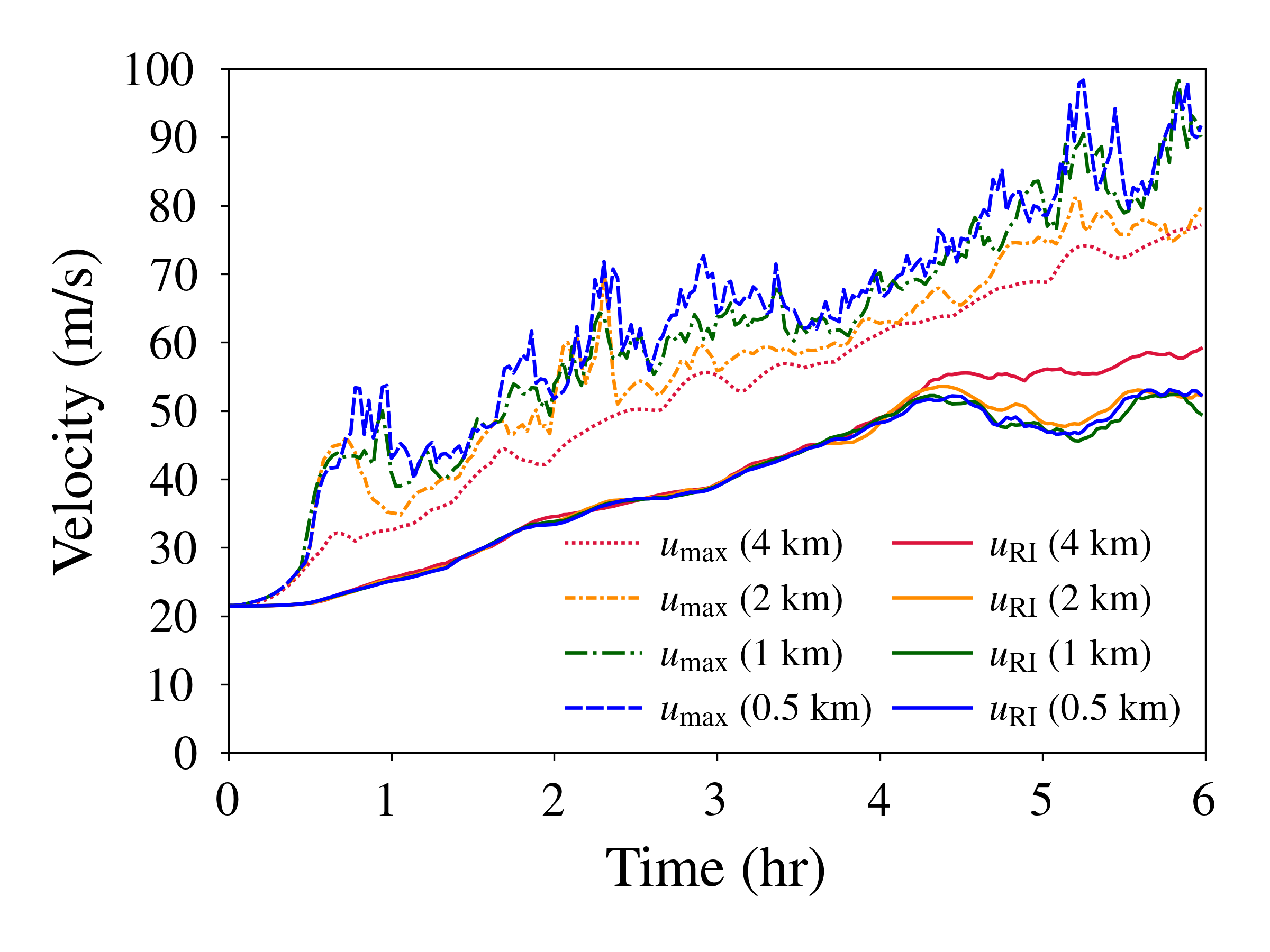}
  \caption{Maximum horizontal velocity ($u_{\mathrm{max}}$) and RI velocity ($u_{\mathrm{RI}}$) at resolutions $\dx=\qty{0.5,1,2,4}$ km.}\label{fig:RIvelocity}
\end{figure}

Figure \ref{fig:RIvelocity} presents the time evolution of the RI velocity at various grid resolutions, compared with the maximum horizontal velocity. The RI velocity is calculated using 50 equi-width strips to average the tangential velocity. At all grid resolutions, the RI velocity exhibits a significant increase exceeding 20 m/s from its initial state. This is higher than the typical RI threshold of 13 m/s, which indicates that the occurrence of the RI event is effectively captured. The RI velocity produces a similar curve across all the resolutions and converges when the grid resolution is higher than $\dx=2$ km, whereas, the maximum horizontal velocity shows larger variance between the profiles of different resolutions.

\subsection{Speedup and energy consumption}

The execution time and energy consumption are compared between a CPU-based simulation using 128 AMD and 4 A100 processors. The test grid consists of 80$\times$80$\times$24 fifth-order elements, which results in 165,888,000 degrees of freedom (DOF). Table \ref{table:energy} summarizes the comparison between the two simulations. The maximum power consumptions are assumed based on the hardware specifications.

Execution times are measured for the time integration loop, i.e., after the simulation setup stage in Figure \ref{fig:porting}, over 100 time steps from the initial state. The CPU run employs the base code, originally optimized for CPUs. The GPU-accelerated code achieves approximately 10.4$\times$ speedup over the CPU code for the current configuration. This speedup is likely due to the higher FLOP capacity of  the GPU and reduced communication overhead from the smaller number of processors. 

The energy consumption per time step in the GPU simulation is approximately 4.84$\times$ lower than that of the CPU simulation. Although a single GPU consumes more power than a single CPU node, the total energy consumption is significantly reduced in the GPU run due to 10.4$\times$ speedup. This indicates that the GPU implementation offers greater energy efficiency compared to the purely CPU-based approach. To emphasize this point,  Table \ref{table:energy} shows that 4 GPUs are 4.84$\times$ more efficient than 128 CPUs; therefore, we require $\frac{128}{4} \times 4.84 = 154$ CPUs to match a single GPU.  To perform equivalent simulations that we performed on 256 GPUs would require nearly 40,000 CPUs (6$\times$ the CPU capacity of Delta).

\begin{table}[h!]
    \caption{Comparison of total energy consumption during 100 time steps between CPUs (AMD EPYC 7763) and GPUs (NVIDIA A100).
    \label{table:energy}}
    \begin{tabular}{lcc}
        \toprule
        & CPU & GPU \\
        \midrule
        Max. Power Consumption (W) & \makecell{ 280 \vspace{-0.4em} \\ \vspace{-0.2em}{\scriptsize (per node)} }  & \makecell{  300 \vspace{-0.4em} \\ \vspace{-0.2em} {\scriptsize (per GPU)} } \\
        Number of Nodes & 2     & 1 \\
        Number of Processors & 128     & 4 \\
        Execution Time Per Step (seconds)  & 5.563 & 0.536 \\
        Energy Consumption Per Step (J)  & 3115.3 & 643.2 \\ 
        \bottomrule
    \end{tabular}
\end{table}
\subsection{Scalability}

To evaluate the scalability of the GPU implementation, we measure the strong and weak scaling performance.

%------------------------------------------------------%
% Strong Scaling
%------------------------------------------------------%

\begin{figure}[h!]
  \setlength{\fboxsep}{0pt}%
  \setlength{\fboxrule}{0pt}%  
  \begin{center}
  \includegraphics[width=0.36\textwidth]{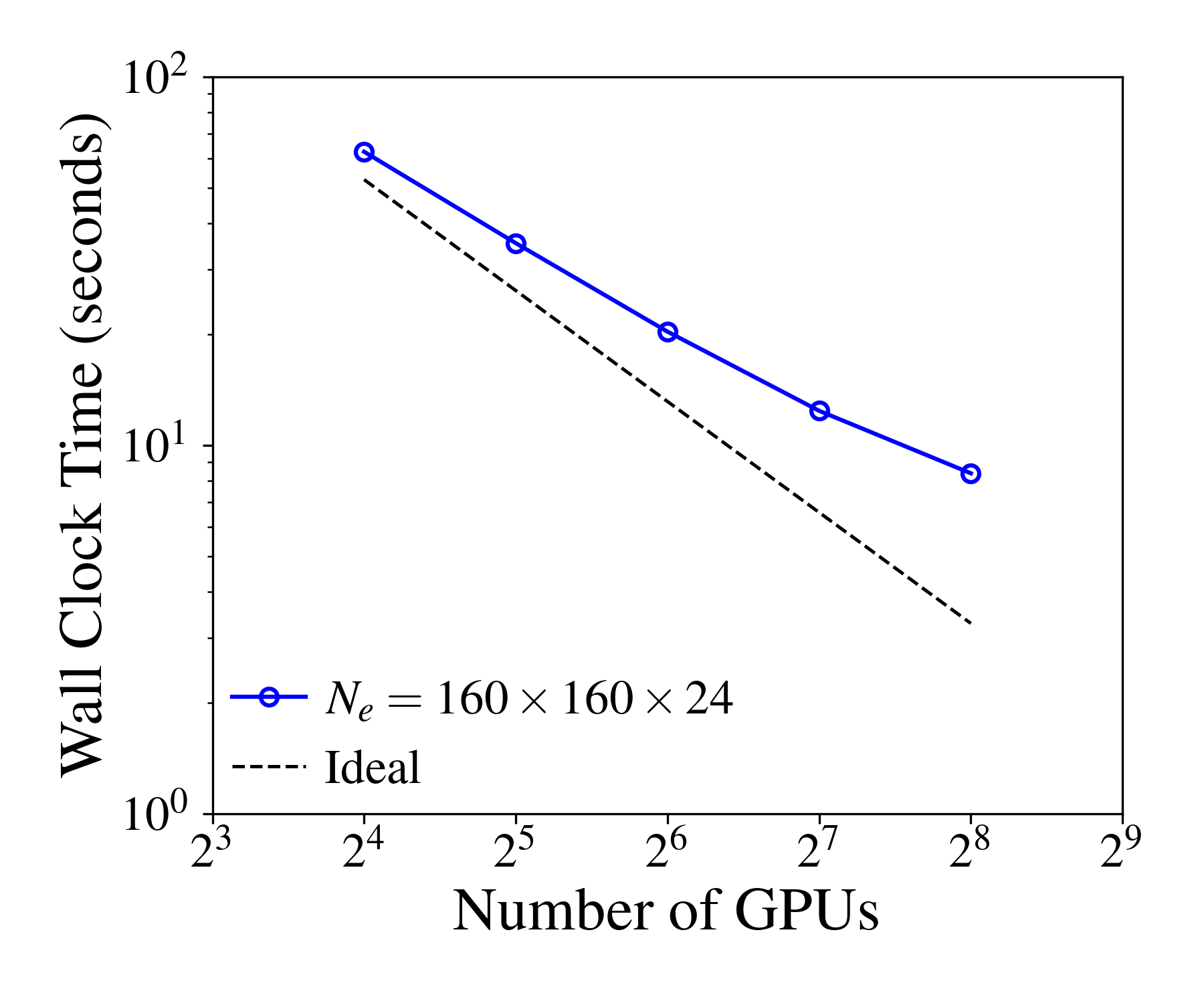}
  \end{center}
  \caption{Strong scaling of the GPU implementation.}\label{fig:strong-scale}
\end{figure}

Figure \ref{fig:strong-scale} shows the strong scaling of the GPU implementation. The grid is composed of 160$\times$160$\times$24 fifth-order elements, which results in a total of 663,552,000 DOF. Strong scalability tends to degrade as the number of GPUs increases. This can be attributed to higher GPU memory access latency, and its effects become relatively more significant as each rank holds a smaller portion of the problem. It is noteworthy that GPUs are the most performant when they are fully saturated.

%------------------------------------------------------%
% Weak Scaling
%------------------------------------------------------%

\begin{figure}[h!]
  \setlength{\fboxsep}{0pt}%
  \setlength{\fboxrule}{0pt}%  
  \begin{center}
  \includegraphics[width=0.36\textwidth]{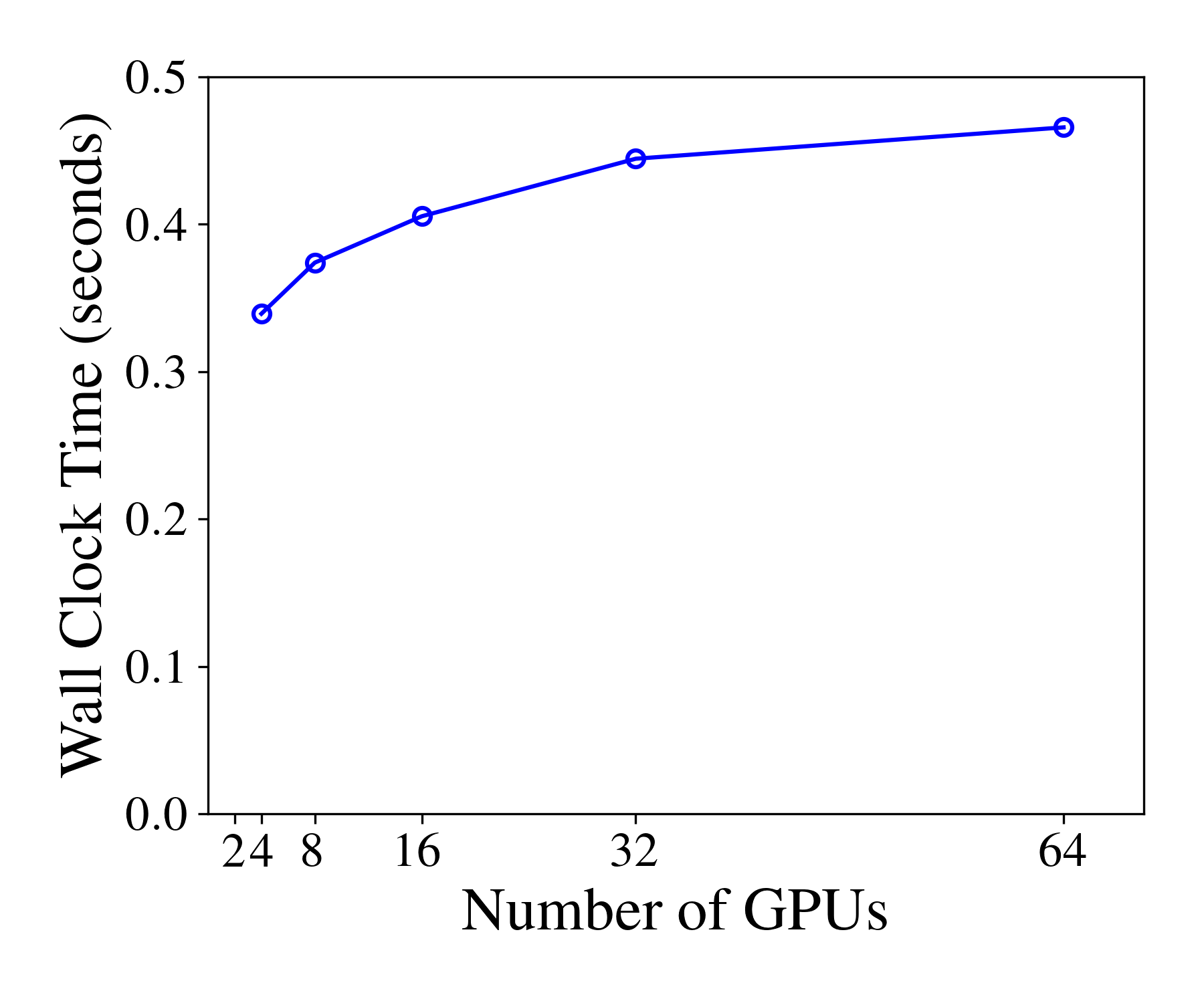}
  \end{center}
  \caption{Weak scaling of the GPU implementation.}\label{fig:weak-scale}
\end{figure}

For weak scaling, the problem size is increased proportionally to the number of GPUs. To achieve this, the grid resolution is doubled while keeping the number of elements per GPU at 40$\times$40$\times$24 (41,472,000 DOF), and the temporal resolution constant at $\Delta t=0.5$ seconds (the largest problem shown has 2.65 billion DOF). The order of basis function is $N=5$ in all direction. Figure \ref{fig:weak-scale} shows that the wall clock time per time step for the weak scaling test. The execution time converges as the number of GPUs increases. This increasing trend is due to more GMRES iterations being required for the implicit solve in larger problem sizes. This result highlights the high performance and scalability of the code.

\subsection{Kernel performance}

We evaluate the performance of the main kernels to compute the RHS vector in Algorithm \ref{alg:rhs} using the roofline model. This model uses the metrics of double-precision floating point operations rates in GFLOPS/s, bandwidth in GB/s, and the arithmetic intensity (GFLOPS/GB). The measured peak performance and bandwidth of NVIDIA A100 GPUs in the test are 7,329 GFLOPS/s and 1,506 GB/s, respectively. 

\begin{figure}[h!]
	\captionsetup[subfigure]{justification=centering}
  \centering
  \begin{subfigure}[b]{0.36\textwidth}
    \includegraphics[width=\textwidth]{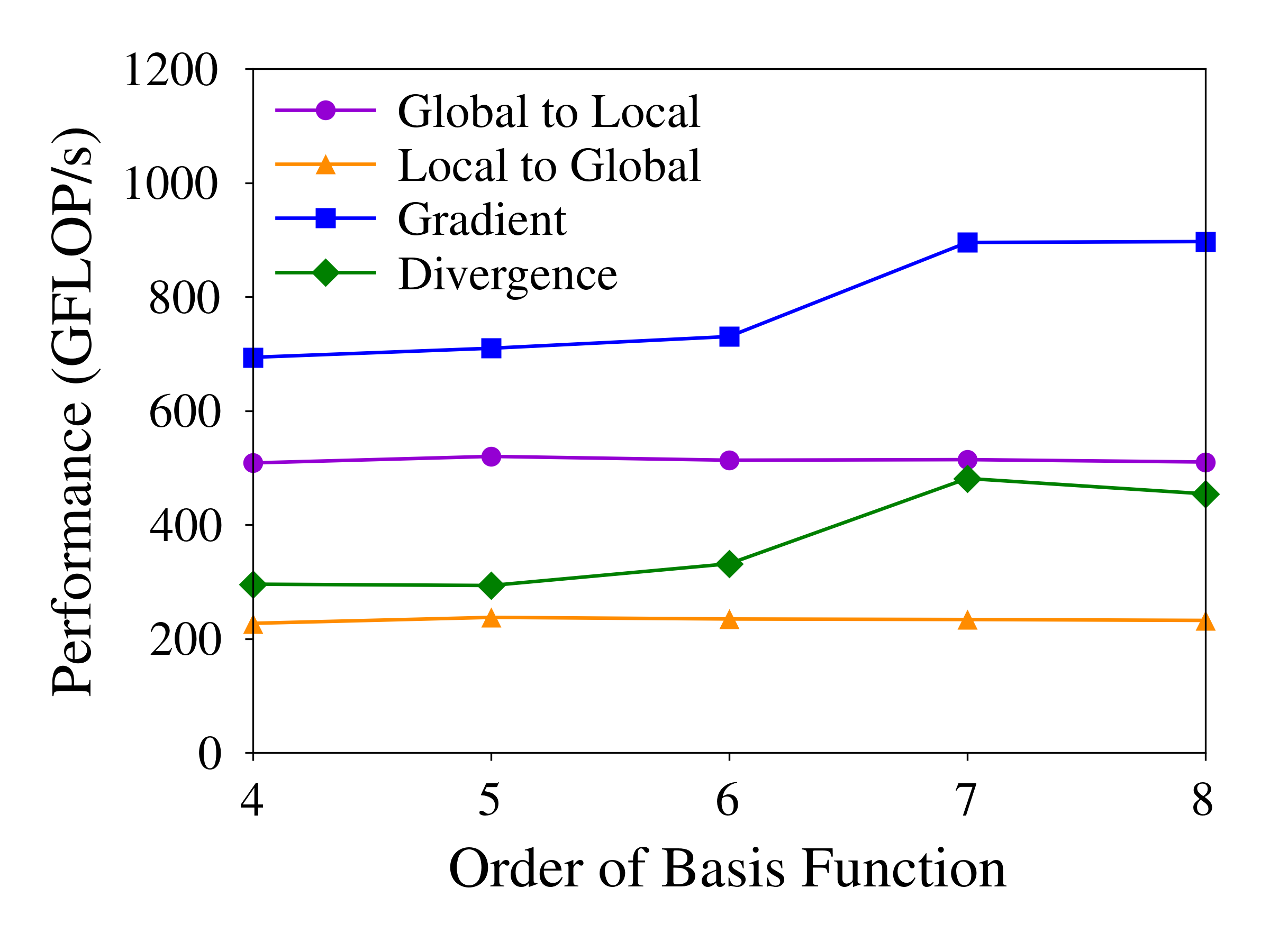}
    \caption{Performance}
  \end{subfigure}
  \begin{subfigure}[b]{0.36\textwidth}
    \includegraphics[width=\textwidth]{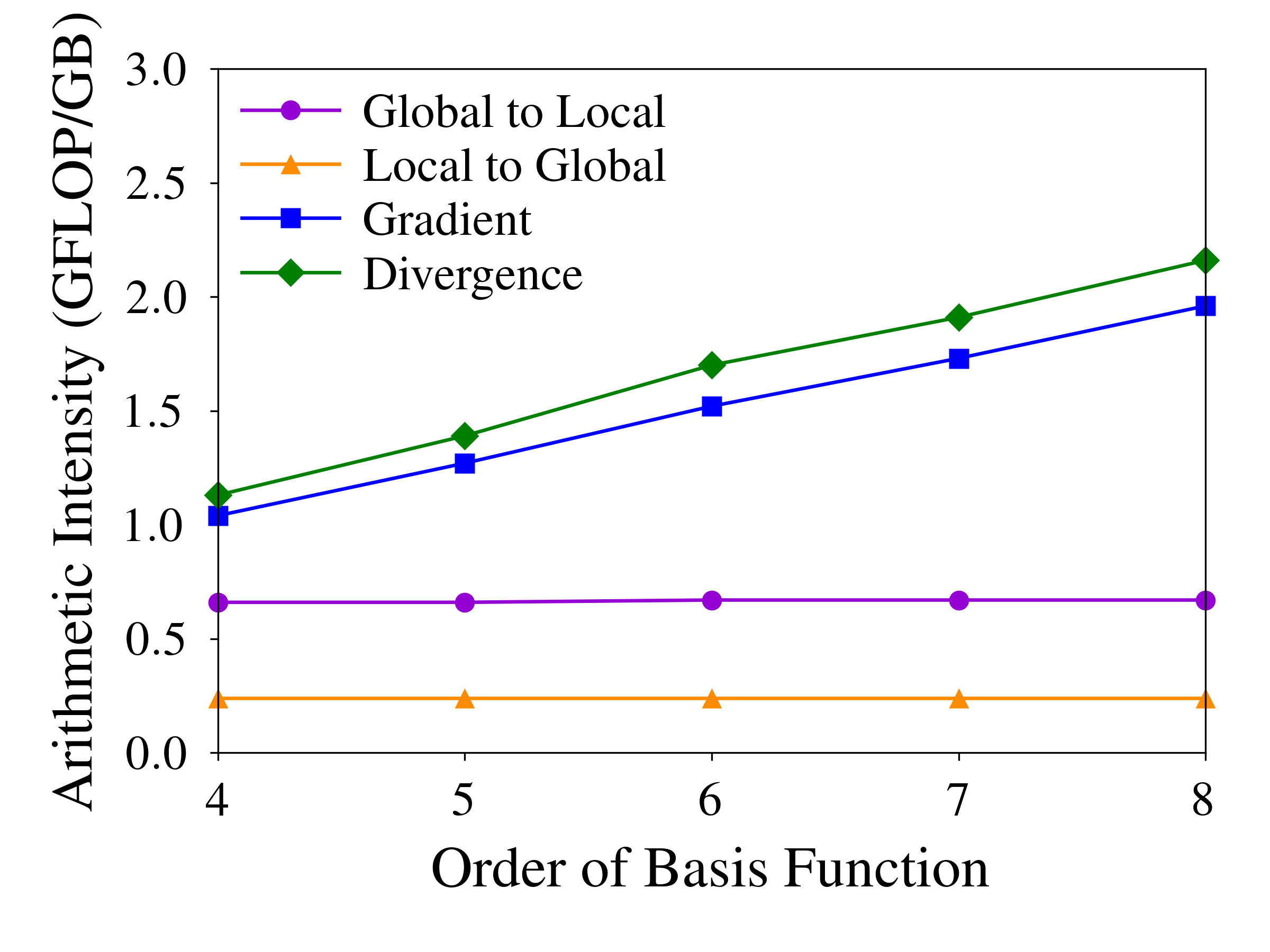}
    \caption{Arithmetic intensity}
  \end{subfigure} 
  \begin{subfigure}[b]{0.36\textwidth}
    \includegraphics[width=\textwidth]{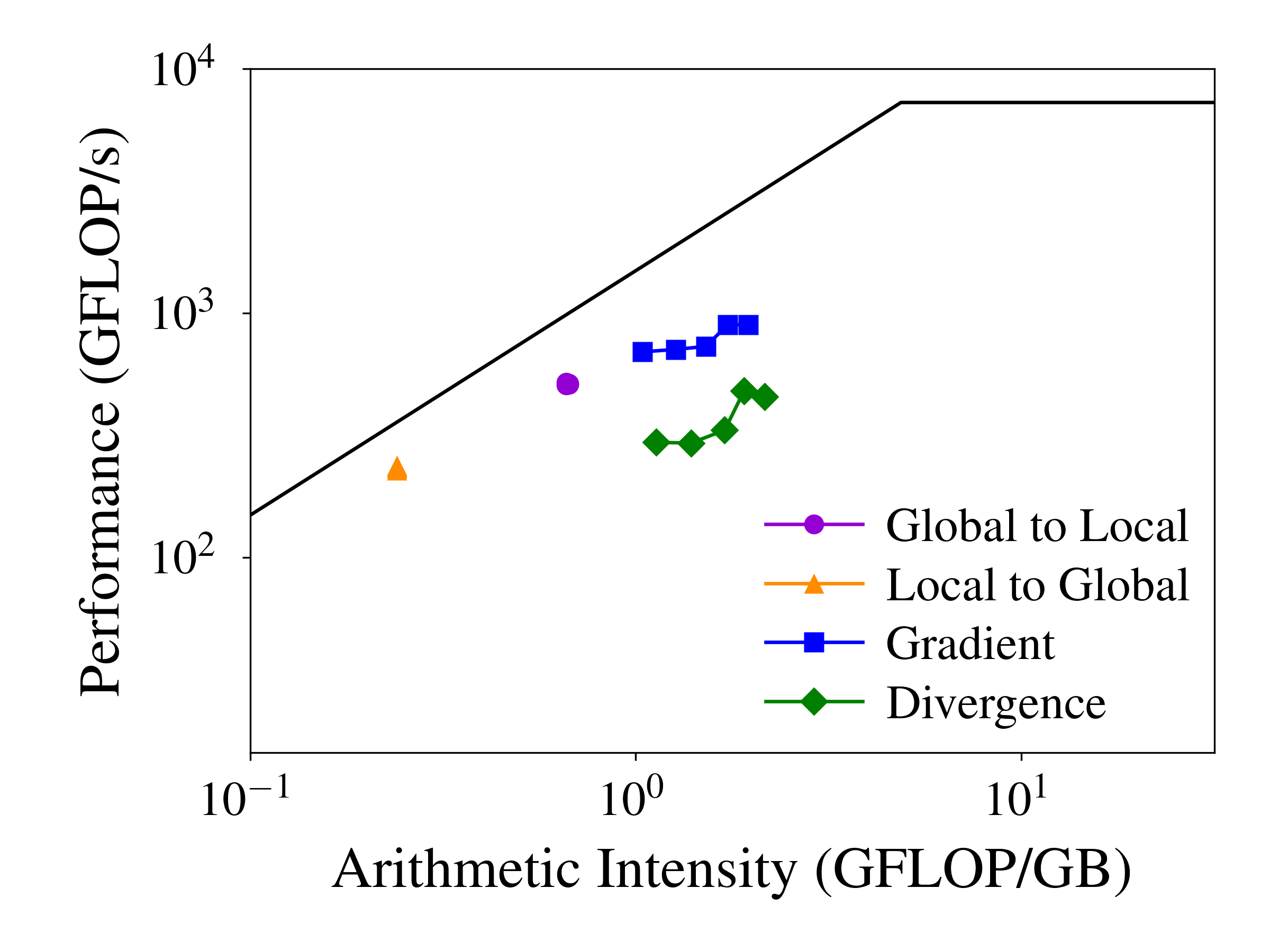}
    \caption{Roofline}
  \end{subfigure}
  \caption{GFLOP/s performance, arithmetic intensity, and roofline model for four main kernels in the RHS function.}\label{fig:roofline}
\end{figure}

Figure \ref{fig:roofline} shows the GFLOP/s performance, arithmetic intensity, and roofline plots for various orders of the basis function. The gradient and divergence kernels are computationally intensive and achieve higher performance as the order of the basis function increases. The measured performance for the gradient kernel ranges from 9.5\% to 12.3\% of the peak performance, while the divergence kernel achieves between 4.1\% and 6.2\%. The lower performance of the divergence kernel compared to the gradient kernel is attributed to the governing equation set given in Eq.~\eqref{eq:governing}, which requires gradient operations for five variables $(p', \ub, \theta)$ whereas the divergence operation is applied only to the momentum, $\rho\ub$. As a result, the divergence kernel involves fewer FLOP computations and becomes slightly more memory-intensive. Higher arithmetic intensity results in higher performance, which implies the potential of high-order methods to better exploit GPUs capabilities. In contrast, the performance of the global-to-local and local-to-global kernels remains independent of the basis function order, 7.1\% and 3.2\% of the peak performance, respectively, as they simply loop over the grid points and their associated DOF.

\section{Conclusions}\label{sec:conclusion}

In this work, we ported the nonhydrostatic atmospheric model xNUMA to GPUs using the OpenACC programming model. This approach facilitates simple maintenance of the code. The entire time-integration process, including implicit and explicit steps, is performed on the GPUs. A matrix-free implementation of the implicit solution is well-suited for GPU computations. 

The model successfully simulates realistic tropical cyclones on GPUs and captures the RI process. The simplified calculation of azimuthally averaged velocity yields converged results of RI across various horizontal grid spacings. 

The GPU-accelerated code yields an energy-efficient simulation, achieving a $4.84\times$ improvement over the CPU version. The GPU implementation demonstrates both strong and weak scaling, as well as efficient kernel performance on NVIDIA A100 GPUs. We expect that this research will help accelerate high-fidelity simulations for numerical weather prediction.

Our future work includes the following:
\begin{itemize}
  \item Using the initial state obtained from observational data to produce more realistic tropical cyclones.
  \item Adding a precipitation microphysics model to simulate moist dynamics of tropical cyclones.
  \item Accelerating the multiscale modeling framework \citep{kang2024multiscale} using GPUs.
\end{itemize}

%------------------------------------------------------%
%   								Acknowledgements
%------------------------------------------------------%

\begin{acks}
  This work was supported by the Office of Naval Research under Grant No.\ N0001419WX00721. F.\ X.\ Giraldo was also supported by the National Science Foundation under grant AGS-1835881. This work was performed when Soonpil Kang held a National Academy of Sciences’ National Research Council Fellowship at the Naval Postgraduate School. Part of Soonpil Kang’s work was performed under the auspices of the U.S.\ Department of Energy by Lawrence Livermore National Laboratory under Contract DE-AC52-07NA27344.\ LLNL-JRNL-2004409. This work used Delta at the National Center for Supercomputing Applications (NCSA) through allocation MTH240030 from the Advanced Cyberinfrastructure Coordination Ecosystem: Services \& Support (ACCESS) program, which is supported by National Science Foundation (NSF).
\end{acks}

%------------------------------------------------------%
%   								Appendix
%------------------------------------------------------%
\appendix

% \input{sections/Complexity.tex}

%------------------------------------------------------%
%   								References
%------------------------------------------------------%

% \begin{thebibliography}{99}
% \bibitem[Kopka and Daly(2003)]{R1}
% Kopka~H and Daly~PW (2003) \textit{A Guide to \LaTeX}, 4th~edn.
% Addison-Wesley.

% \bibitem[Lamport(1994)]{R2}
% Lamport~L (1994) \textit{\LaTeX: a Document Preparation System},
% 2nd~edn. Addison-Wesley.

% \bibitem[Mittelbach and Goossens(2004)]{R3}
% Mittelbach~F and Goossens~M (2004) \textit{The \LaTeX\ Companion},
% 2nd~edn. Addison-Wesley.

% \end{thebibliography}

\bibliography{references}
\bibliographystyle{SageH}

\end{document}